\documentclass[useAMS,usenatbib]{mn2e}

\usepackage{graphicx}
\newcommand{\putfig}[2]{\includegraphics[width=#2]{#1.eps}}

\usepackage{graphics}
\usepackage{rotating}
\usepackage{amsmath}
\usepackage{amssymb}

\def\Nfields{8,385\,}
\def\Nsources{9,040\,}
\def\F0{$\mathcal{F}_0$}
\def\Fda{$\mathcal{F}_{\Delta\alpha}$}
\def\PMNCA{PMN-ATCA}

\title[ATPMN: \Nfields PMN sources]{ATPMN: accurate positions and flux densities at 5 and 8~GHz for \Nfields sources from the PMN survey}
\author[D. McConnell et al.]
  {D.~McConnell,$^1$ E.M.~Sadler,$^2$ T.~Murphy,$^{2,3}$R.D.~Ekers$^1$\\
  $^1$CSIRO Astronomy \& Space Science, Australia Telescope National Facility, P.O. Box 76, Epping, NSW 1710, Australia\\
  $^2$Sydney Institute for Astronomy, School of Physics, The University of Sydney, NSW 2006, Australia\\
  $^3$School of Information Technologies, The University of Sydney, NSW 2006, Australia}

\begin{document}
\pagerange{\pageref{firstpage}--\pageref{lastpage}} \pubyear{2011}
\maketitle
\label{firstpage}

\begin{abstract}
We present a source catalogue of \Nsources radio sources resulting from  high-resolution observations of \Nfields PMN sources with the Australia Telescope Compact Array.  The catalogue lists flux density and structural measurements at 4.8 and 8.6~GHz, derived from observations of all PMN sources in the declination range  $-87\degr < \delta < -38\fdg5$ (exclusive of galactic latitudes $|b| < 2\degr$)  with flux density $S_{4850}  \geq 70$~mJy (50~mJy south of $\delta = -73\degr$).  We assess the quality of the data, which was gathered in 1992--1994,  describe the population of catalogued sources, and compare it to samples from complementary catalogues.  In particular we find 127 radio sources with probable association with $\gamma-$ray sources observed by the orbiting Fermi Large Area Telescope.  
\end{abstract}


\section{Introduction}

Radio surveys of the sky contribute to the body of multi-wavelength data used in astrophysics for understanding the Universe.  The time required for radio surveys increases with sensitivity, resolution and observing frequency. In particular, wide-area surveys at cm-wavelengths that can yield resolutions comparable to those of optical images require very long observations and are often only feasible with special techniques.  The entire sky has been surveyed with a resolution of about 45 arcsconds with the VLA in the north at 1.4~GHz (NVSS, \citealt{1998AJ....115.1693C}) and with the MOST in the south at 0.843~GHz (SUMSS, \citealt{Mauch:2003ed}).  At higher frequencies the telescope field of view shrinks and the time required to survey a given area of sky increases as the second power of frequency.  This law was overcome at 5~GHz through the use of a seven-beam receiver on the Green Bank 91m and  Parkes 64m telescopes for northern \citep{Condon:1989bt} and Parkes-MIT-NRAO (PMN) survey of the southern sky \citep{Griffith:1993cl,Wright:1994gh,Wright:1996iq,Griffith:1994fz,Griffith:1995ep}, but at the cost of the lower resolution of 3--5 arcminutes. 

Following the PMN survey, the Australia Telescope Compact Array (ATCA) was used to re-observe a sample of \Nfields sources in the PMN catalogue \citep{Wright:1997wz}, and so measure the sources with high resolution without needing to survey the whole sky.  By observing the selected 2.5 per cent of the sky within the ATCA survey's declination limits, 37 per cent of the PMN sources in that area were characterised with high resolution.  This paper describes the re-analysis of these observations and presents the Australia Telescope-PMN (ATPMN) catalogue of radio sources in the southern sky.

At 20~GHz the ATCA was used in a novel mode to enable a rapid wide-area survey \citep{Murphy:2010ej}.  A three-element array was formed with a broad-band analogue correlator and the sky was scanned rapidly along the southern meridian \citep{Hancock:2011fb}.  From these scans, fields containing detected sources were selected for re-observation with the ATCA in its conventional mode, and characterised at 20, 8 and 5~GHz.

In Table \ref{surveys} we give the parameters of a number of catalogues of southern radio sources, including those mentioned above.  We include in the list CRATES \citep{Healey:2007fa}, a compendium of flat-spectrum sources compiled from several catalogues and observing campaigns.
\begin{table*}
 \centering
 \begin{minipage}{140mm}

 \caption{Southern radio source samples.}
\begin{tabular}{lccccrc}
  \hline
 \multicolumn{1}{c}{Sample}  & Frequency & Resolution & Sky & flux density limit &  \multicolumn{1}{c}{Sources} & Reference\\
 \multicolumn{1}{c}{name}     & (GHz)        &    (arcmin) &  limits     &  (mJy)              & \\
\hline
MRC      & 0.408      & 2.6          & $-85\fdg0 < \delta < +18\fdg5$   & 700 & 12,141 & \cite{Large:1981vm}\\
PMN      & 4.85      & 4.2          & $\delta < +10\degr$   & 20--45 & 50,814 & \cite{Griffith:1993cl}\\
SUMSS   & 0.843    & 0.75        & $\delta < -30\degr   $ & 8 -- 18        & 211,063 & \cite{Mauch:2003ed}\\
AT20G   & 20, 8, 5 & 0.17        & $\delta < 0\degr$       & 40\footnote{At 20~GHz} & 5,890  &  \cite{Murphy:2010ej}\\
CRATES  & 8.4        & 0.17,0.02,0.003\footnote{From two configurations of the ATCA, and the VLA A configuration}          & all sky                           & 50 & 10,369 & \cite{Healey:2007fa} \\
ATPMN  & 4.8, 8.6 & 0.03,0.02 & $\delta < -38.5\degr$ &     7\footnote{Incomplete below the 70mJy PMN selection level}    & \Nsources &  this work\\
\hline
\end{tabular}
 \label{surveys}
\end{minipage}
\end{table*}

The Parkes-MIT-NRAO (PMN) radio survey of the sky south of declination +10\degr was conducted with the Parkes 64-m radiotelescope at 4850~MHz in June and November of 1990.  It was conducted in four declination zones: Southern ($-87\fdg5 < \delta < -37\degr$), Zenith  ($-37\degr < \delta < -29\degr$), Tropical  ($-29\degr < \delta < -9\fdg5$) and Equatorial  ($-9\fdg5 < \delta < +10\degr$)
\citep{Griffith:1993cl,Wright:1994gh,Wright:1996iq,Griffith:1994fz,Griffith:1995ep}.  The PMN survey  complemented and extended the surveys of the northern sky by \cite{Condon:1989bt}.  It had a declination-dependent lower flux density limit of 20--72~mJy and resulted in a catalogue of 50,814 sources in the declination range +10\degr to $-$87\degr.  It had spatial resolution (full width at half-maximum: FWHM) of 4\farcm2 and a flux density--dependent positional accuracy of 10\arcsec to 60\arcsec.  \cite{Gregory:1994bx}  published an alternate analysis of the Southern PMN survey.

From 1992 November to 1994 March, observations of  \Nfields fields centred on a selection of PMN sources south of declination $-38\fdg5$ were made at 4.8 and 8.6~GHz with the  ATCA.  The primary motivation for the ATCA follow-up survey was to provide source positions precise enough for unambiguous optical identification. In addition, the survey would provide structural and polarization information for each source, and an estimate of the spectral energy distribution in the cm-wavelength range, significantly amplifying the value of the PMN survey itself.

The observations, data reduction and the resulting catalogue (referred to here as \PMNCA) are described in an unpublished report by \cite{Wright:1997wz}.  \citeauthor{Wright:1997wz} described the analysis procedures for the ATCA data, procedures chosen for speed of processing at the expense of completeness and reliable characterisation of complex fields.  The resulting \PMNCA\ catalogue has been available online since the date of that report.   Since the time of the initial data reduction, advances have been made in both the facilities available for ATCA data reduction, and in the understanding of the performance of the ATCA itself.  Furthermore, we have now significantly improved flux density and position information on the reference sources used for ATCA data calibration.  These factors, and the ready availability of the original data from the Australia Telescope Online Archive (ATOA\footnote{The ATOA is available at {\tt http://atoa.atnf.csiro.au}}) have motivated us to re-analyse the survey data.  Here we describe that re-analysis and present a new catalogue, the ATPMN catalogue, of \Nsources sources. 

In Section 2 we describe the original field selection and observations; in Section 3 we describe the new data reduction process and discuss data quality and the factors that affect the reliability of the results.  Section 4 introduces the catalogue.  In Section 5 we give an overview of the source population in the catalogue, make some comparisons with other catalogues of sources in the south (including the unpublished \PMNCA\ catalogue derived from the same data), and briefly discuss some aspects of the dataset that will be explored more thoroughly in future work.

\begin{figure*}
\begin{center}
\includegraphics[width=\textwidth]{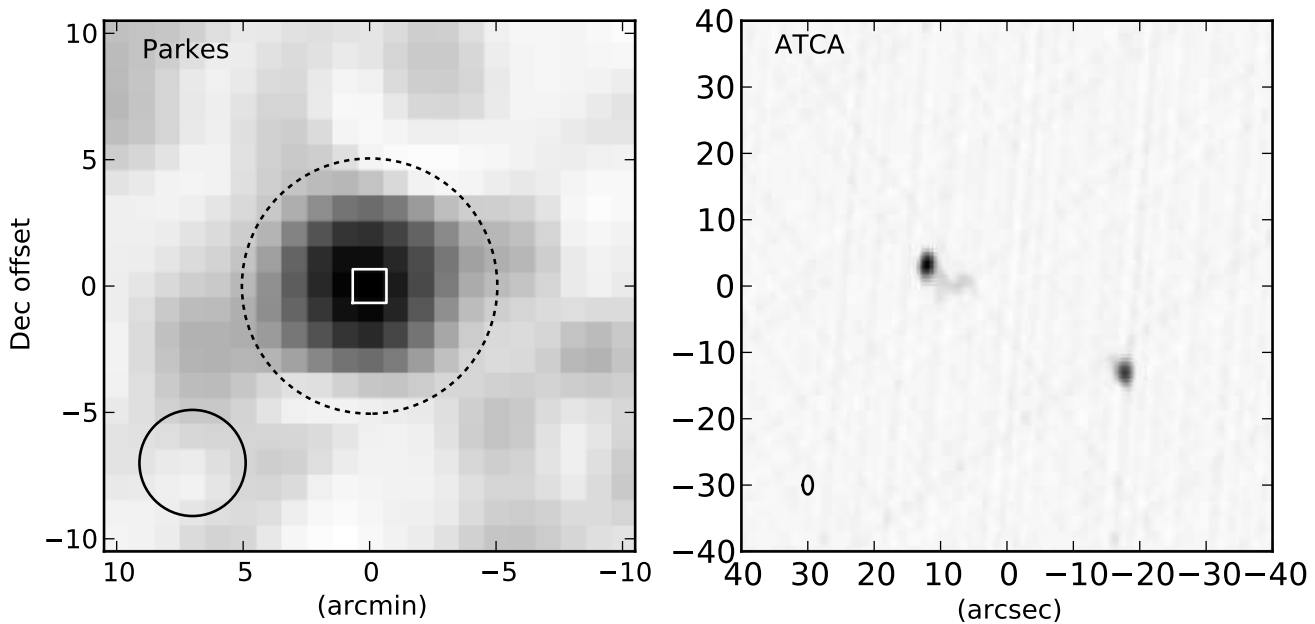}
\caption{Images of the source PMNJ0010$-$4422 ($\alpha = $~00:10:30.4, $\delta = -$44:22:53.0)  with $S_{4850} = 369$~mJy.  On the left is the Parkes 4.85~GHz image with its 4\farcm2 beam (FWHM) shown at lower left.  On the right is the ATCA 4.8~GHz image with a $2\farcs8 \times 1\farcs$5 synthesized beam.  The dotted circle and the small white square on the Parkes image indicate the ATCA field of view (to half power) and the extent of the image displayed on the right. The ATCA primary beam sizes are 10\farcm1 and 5\farcm9  at 4.8 and 8.6~GHz.  The two components visible in the ATCA image have a total flux density of 261~mJy.} 
\label{examplefield}
\end{center}
\end{figure*}

Throughout this paper we refer to the positions selected from the PMN catalogue to be re-observed with the ATCA as ``fields'', or ``PMN sources'' when the properties of the source are relevant.  We refer to the sources identified from the ATCA observations as ``sources''.  Frequently a single field contains several sources; that is a PMN source, when observed with the ATCA is resolved into several components that we refer to as sources. Fig. \ref{examplefield} illustrates such a case.

The motivation for this work is the publication of improved knowledge of the brightest southern PMN sources: accurate positions and structural information at the PMN frequency of 4.8~GHz and at 8.64~GHz, and flux densities at both frequencies.  

\section{Observations and initial data analysis}
The Compact Array was used to re-observe the strongest sources contained in the Southern PMN survey \citep{Wright:1997wz}.  Sources were chosen from the declination range $-87\degr < \delta < -38\fdg5$, and limited to galactic latitudes $|b| > 2\degr$ to avoid the field complexity expected in the Galactic plane.   The total sky area within these limits is 2.25 steradians, about 18 per cent of the sky. Within these limits, all PMN sources with 4850~MHz flux density $S_{4850}  \geq 70$~mJy (50~mJy south of $\delta = -73\degr$) were included to form  the final sample of \Nfields to be included in the ATCA survey.

The ATCA \citep{1992JEEEA..12..103F} has six 22-m moveable antennas and for the observations described here was used in several of its longest, 6~km arrays.  The ATCA observations were made over the period 1992 November to 1994 March in five separate sessions.  During each session the ATCA was configured in one of the standard 6~km configurations.  Table~\ref{sessions} gives the dates and Array configuration used for each set of observations. Observations were made simultaneously in two bands centred at 4.8 and 8.64~GHz.  Each band had a useable width of 100~MHz.  ATCA antennas are sensitive to orthogonal linear polarizations; all correlation products were recorded to allow measurements in all four Stokes parameters.

\begin{table}
 \centering
 \caption{The ATCA observations were made by  \protect\cite{Wright:1997wz} in the five separate sessions listed here.  For the re-analysis described here we have recovered the original data from the ATNF online archive
(ATOA).}

\begin{tabular}{clcc}
  \hline
   Epoch     &  \multicolumn{1}{c}{Dates}    & Array & Shortest \\
                  &                                              &          & spacing (m)\\
\hline
1 & 1992 Nov 09--15 & 6A & 337\\
2 & 1993 Mar 03--07 & 6D & 76 \\
3 & 1993 Jun 11--15 & 6A & 337 \\
4 & 1993 Sep 23--26 & 6D & 76 \\
5 &1994 Mar 11--14 & 6C & 153 \\
\hline
\end{tabular}
 \label{sessions}
\end{table}

Three types of fields were observed: the primary calibrator, the secondary calibrators and the target fields.  The primary calibrator source PKS~B1934$-$638 was used to determine corrections for the instrumental spectral and polarization response.

The secondary calibrators were chosen to be point sources with known positions and are used for determining the time-dependent parts of the instrumental gain.  In particular, the fluctuations in visibility phase introduced by the atmosphere are estimated with the secondary calibrator observations.  Over the entire set of observations, 114 different secondary calibrators were observed, 57 of them in more than one observing session.   The targets were the \Nfields fields centred on the subset of PMN sources selected as described above.

Each target field was observed three times at hour angles sufficiently spaced to allow the characterisation of sources in simple fields.  Each observation was of 30-second duration, giving a point source sensitivity of about 1~mJy.  In practice, this sensitivity was not always achieved; poor seeing (atmosphere induced phase fluctuations) and field complexity and confusion could both degrade the performance.  For each twelve-hour observing period, a declination band was chosen so that all target fields within that band could be observed three times -- at hour angles near $-4^h$, $0^h$ and $+4^h$.  Secondary calibrators were observed for two minutes at intervals of approximately 30 minutes at hour angles close to those of the adjacent target fields.  At times instrumental or operational problems lead to the omission of one or more observations of a given target field.  Such fields were scheduled for re-observation on a later date.  In all, 502 target fields were observed more than once. The primary calibrator was observed at least once per day for about five minutes.

\cite{Wright:1997wz} described the analysis procedures used to produce the original source catalogue.  The analysis proceeded in two stages: each 30-second scan was reduced with 1-dimensional imaging and deconvolution to a simple 3-parameter description of up to two sources; once results were available for all scans of a given field, the 2-dimensional source information for that field was derived from the individual 1-dimensional descriptions.  It was recognised that this approach was less able to characterise complex fields than the conventional technique that optimally combines all data for each field to produce 2-dimensional images.  The 1-dimensional approach was chosen to be economical in both computing time and data-storage, both of which were more expensive than they are today. \citeauthor{Wright:1997wz} reduced a small sample of ATCA fields using the conventional methods to estimate the likely increase in information obtainable from some future complete re-analysis of the dataset. The  \citeauthor{Wright:1997wz}  catalogue was not published, but has been available online since its production, and has been used in other works (for example \cite{Healey:2007fa}).  Our work has aimed to improve on the original analysis by using a full 2-dimensional approach for all fields, and to benefit from improved knowledge of the ATCA's performance and its astrometric calibration.  The remainder of this paper concerns that re-analysis.

\section{Data processing}
We now proceed to describe a full re-analysis of the ATCA follow-up of the selected PMN sources.
We processed the data  in two phases: calibration and analysis.  Both phases relied on the \textsc{miriad} software package \citep{Sault:1995ub} which we used through  a series of scripts to ensure consistent treatment of all fields in both calibration and analysis.

\subsection{Calibration}
\label{calibration}
Each target field was assigned a secondary calibrator,  chosen for proximity on the sky and in time of observation.  A table of gain corrections for each secondary calibrator were built up as follows: bandpass and polarization corrections were copied from the results of the calibration of the instrument using the primary calibrator; the relative variations in gain were then deduced from the secondary calibrator data themselves; finally the gains were scaled so as to put the secondaries onto the flux density scale of the primary calibrator.  Target field data were corrected by application of the gain table of its assigned secondary calibrator.

At this stage in the process, two other steps were necessary to account for some shortcomings in the original choice and observation of some of the chosen secondary calibrators. A significant fraction of them were scheduled for observation at the wrong position.  That is, they were not observed at the phase reference centre of the array.\footnote{During the early years of ATCA operation the characteristics of calibration sources were measured, and the errors in calibrator positions was a consequence of the immaturity of the calibrator catalogue: accuracy of many of the calibrator positions has improved since 1992.}  Unless corrected, this leads to a systematic shift in phase corrections applied to target field data and the position error being propagated to source positions determined in target fields.  We amended the calibration routine  to include the correction of this error in the secondary calibrator data using recent calibrator measurements, prior to the determination of the time-dependent gain corrections.  The second extra step in processing was necessary for a small subset of sources chosen as calibrators (B0407$-$658, B0939$-$60, B1329$-$665 and B1756$-$663) that are resolved at 4.8~GHz and 8.6~GHz on baselines shorter than 6~km.  Using the standard calibration for these sources would result in incorrect amplitude corrections being passed on to target field data.  The amended calibration routine used for these sources solved only for the time-dependent phase, and relied on the amplitude corrections derived from primary calibration.  Although the sources listed above were not suitable for calibrating these data, they remain in the ATCA calibrator catalogue and can be used effectively for more compact configurations  of the ATCA.

Two calibration solutions were determined for the secondary calibrators.  The first used the shortest possible integration time to avoid depression of the measured amplitude (flux-density) of the source due to atmospheric phase fluctuations.  This calibration was used when doing flux-scale comparisons described later.  The second used integration times to better represent phase variations experienced by the target sources, and was used for their calibration.

\subsubsection{Phase calibration in two stages}
\label{phasecalibration}
With sufficient signal, self-calibration of ATCA data can significantly improve the calibration of gain phases.  Just as corrections to the antenna gains are derived from observations of point calibration sources, observations of the target sources themselves can be used provided some knowledge of the source structure.  Many of the PMN fields contain easily-modelled sources and good phase corrections can be determined from their visibility measurements.  The observing pattern used for the ATPMN survey allows us to use corrections for those fields across the whole data set.  Secondary calibrators were observed at intervals of about 30 minutes, bracketing a number of target fields observed at approximately the same position in the sky.  We have found that fields with peak brightness exceeding about 40 mJy/beam, about half of all fields, yield reliable phase corrections from self-calibration.   For gain phases, the full calibration has the following steps:
\begin{itemize}
\item apply phase corrections to all target fields as interpolations of corrections determined from secondary calibrator observations;
\item derive phase corrections using  \textsc{miriad} task \textsc{gpscal}  from all fields satisfying the peak brightness criterion; if the criterion was satisfied at 4.8~GHz only, the 8.6~GHz corrections were taken from the 4.8~GHz solution;
\item apply phase corrections to all target fields as interpolations of the self-calibration results.
\end{itemize}
In this way, all fields have either self-calibration, or phase corrections interpolated over much shorter intervals than is possible with secondary calibrators alone.  Note that self-calibration breaks the phase reference with the secondary calibrators and so can lead to a loss of accuracy in absolute position.  Our model-fitting procedure described below includes a step to avoid such a loss.

\subsection{Imaging and source fitting}
\label{imagingandfitting}
We have estimated the position, angular size and flux density of all sources detected in each field,  taking advantage of the simultaneous measurements in the 4.8 and 8.6~GHz bands.  Our analysis recognises the possibility of a pair of close sources being incompletely resolved at the longer wavelength and appearing as one while being clearly resolved into a pair at the shorter wavelength.  It also recognise the possibility of a source visible in one of the wavebands falling below the nominal detection threshold in the other. Analysis proceeded in the following steps.
\begin{enumerate}
\item Image formation : Form a cleaned  total intensity (Stokes $I$) image from the calibrated visibility data. 
\item Source finding : Compile a list of all sources in the image with peak brightness above $5 \sigma$.
\item Double detection : Take special care to recognise pairs of sources close enough to appear as one to the source finding software.
\item Initial model fitting : Refine the parameters of all sources in each field with a least-squares fit to the visibility data.
\item Reconciling the two-frequency measurements : Form a list of distinct sources in each field, taking account of the two sets of measured components, one for each frequency band.
\item Interactive component selection : Intervene in the component selection for the small fraction of fields for which the fitting procedures above fail.
\item Final model fitting  : Refit the final list of sources to the visibility data in both wave bands and in Stokes $I$, $Q$, $U$ and $V$ components. 
\end{enumerate}

We now describe these in detail.
  
\subsubsection{Image formation}
For each field, we formed  images at 4.8~GHz and 8.6~GHz from the calibrated, uniformly weighted visibility data.  At these frequencies, the sizes of the ATCA primary beam are 10.2 arcminutes and 5.9 arcminutes, and typical synthesized beams are 2.2~$\times$~1.6~arcseconds and 1.3~$\times$~0.9~arcseconds (all expressed as full width at half power).  We formed images of  2048~$\times$~2048 pixels of 0.5 and 0.3 arcseconds respectively, giving image sizes of 17.1 and 10.2 arcminutes that extend to the 10 per cent point of the primary beam.  In each case the whole image was deconvolved using the \textsc{miriad} task \textsc{clean}  which uses a hybrid Hogb\"{o}m/Clark/Steer Clean algorithm.

\subsubsection{Source finding}
We used the  \textsc{miriad} task  \textsc{imsad} to locate sources in the each image, specifying a detection threshold of  5$\sigma_{im}$, the $rms$ brightness across the image . Each source is characterised by its position, flux density and the ellipse defining the half-power points of the gaussian brightness distribution.  We retain only the flux density and position parameters for those sources with gaussian shape parameters that are consistent with (no smaller than) the synthesized beam.

\subsubsection{Double detection}
The procedure described above can fail to discriminate between two close sources  when the region between the sources is brighter than the detection threshold.  In such cases, a single gaussian fit is returned for the region encompassing both sources.  To recover more information about such close pairs of sources,  we took the following steps for any source whose major axis exceeds twice the size of the synthesized beam:
\begin{enumerate}
\item define a rectangular region in the image large enough to include the source in question;
\item determine the maximum brightness $B_{max}$ in this region;
\item use  \textsc{imsad} to look for sources above eight brightness thresholds logarithmically spaced between 5$\sigma_{im}$ and $B_{max}$;
\item keep the parameters for two sources if detected, otherwise retain the original single parameter set for this region.
\end{enumerate}

\subsubsection{Initial model fitting}
\label{modelfitting}
Using the  \textsc{miriad} task  \textsc{uvfit}, we refined the description of candidates sources by means of a least-squares fit to the visibility data.  This step accepts up to four candidate sources, more than  \textsc{uvfit} can process in an unconstrained fit.  If necessary, a preliminary constrained fit is performed for which the estimated position is held fixed; the results are ranked by flux and the brightest three sources are used in the main, unconstrained fit.  The results are held as two lists of source parameters, one for each frequency band: we call each parameter-set a ``measurement''.  In rare cases  \textsc{uvfit} fails to converge because of being under-constrained by the data.  In such cases our procedure is to fix one coordinate (declination) of each source, beginning with the least significant, until a stable solution is found.  If this strategy also fails, sources are dropped from the candidate list, weakest first, until a solution is found.  In a small number of cases this results in no sources being characterised for a field.

\subsubsection{Reconciling the two-frequency measurements}
At this stage of the analysis we have two source lists for each field, one for each observing band.  From these we form a single list of sources,  registering two sources where ever a single source in the lower resolution image (at 4.8~GHz) overlapping a pair of sources that appear distinct at 8.6~GHz.  This source list forms the input for the next phase of model fitting.  After each fit we reassess the components of the refined model and remove implausible or low-significance components from the model.  This stage concludes when the number of components in the model stabilises.  We have developed, from experience, a set of criteria to judge whether components should be rejected.  We reject any that satisfy the following relations:


\begin{enumerate}
\item source flux density is below the limit A $S < S_A$;
\item source is overextended: significant flux on less than four baselines;
\item source has flux density much less than the strongest source in the field: $S < 0.05\times S_{max}$;
\item source flux density is below the limit B $S < S_B$ and $S < 0.1\times S_{max}$;
\item $S < S_B$ and $S < S_{max}$ and the source is not detected in the other frequency band;
\item source lies on a radial side-lobe of the strongest source in the field and is not detected in the other frequency band;
\item source is detected at 8.6~GHz, but not at 4.8~GHz.
\end{enumerate}

The values of $S_A = 7$~mJy and $S_B = 20$\, used were chosen after some experimentation, as a compromise between including too many random fluctuations or artifacts of impefect calibration and rejecting real sources. Note that $S_A$ is approximately five times $\sigma_{im}$, the image $rms$ for most fields. 

\subsubsection{Interactive component selection}
The heuristic procedure described above was developed to analyse the range of fields encountered in the PMN source sample.  However, it failed on a small fraction (about 0.4 per cent) of fields. These fields were analysed interactively using \textsc{uvfit}, giving more control over the set of candidate sources used for the fit.

\subsubsection{Final model fitting}
The analysis of each field is concluded by forming a final set of measures of each component in both frequency bands.  This step included a fit constrained to determine source position only from the visibility data {\em without} the phase corrections determined by self-calibration (see \textsection\ref{phasecalibration}).  Estimates of Stokes quantities $Q$, $U$ and $V$ are made using \textsc{uvfit}, holding source position and size fixed and fitting only for flux density.  Each measure includes estimates of uncertainties in all measured quantities, recorded as the formal numerical errors reported by \textsc{uvfit}.  Each measure also includes a set of flags indicating certain conditions that may influence the reliability of the measurement, conditions that in some cases are reflected in the uncertainty estimates.

\subsection{Data quality and reliability of the results}
\label{dataquality}
For each quantity characterising sources in this catalogue we have estimated a likely measurement error, and for each source we have assigned a set of quality flags that reflect cases with large measurement errors, or indicate the possibility of some systematic error unaccounted for by the quoted errors.  In this section we describe the main sources of error in the final results; we present a characterisation of overall data quality; and for some sources we compare the new results with other measurements of the same object.

In this work source properties are determined from just three brief observations giving a sparse sampling of the $uv$--plane.  A small amount of bad or missing data can severely degrade the quality of the results.   The main threats to quality in these data and the reliability of our results were atmosphere--induced phase fluctuations, instrumental failures, and complexity of the observed fields.  For some fields, insufficient good data were recorded to properly form any results. Of the \Nfields fields observed, 166 could not be processed for lack of data.  

\subsubsection{Atmospheric effects}
The observing program included frequent short observations of known point sources -- the secondary calibrators.  Measurements of these allow the correction of fluctuations in both amplitude and phase, provided the fluctuations are not too big, and slow relative to the 30-minute calibrator interval.  Amplitude variations for the ATCA are typically small and occur over times longer than the calibration cycle.  Phase variations are induced by the inhomogeneous atmosphere; the size of phase variations can differ markedly from day to day.  For these observations, three 30-second integrations separated by about four hours, the phase variations have two distinct effects.  Phase variation during the brief integration of each scan causes decorrelation and a reduction of the observed visibility amplitude.  Phase variations on the longer time scale also cause decorrelation and in severe cases, can degrade the image to the point that false sources may be identified.  The second stage of phase calibration, using self-calibration of suitable target fields, is necessary to successfully track phase variations from scan to scan, and in most cases is successful.

\begin{figure}
\begin{center}
\putfig{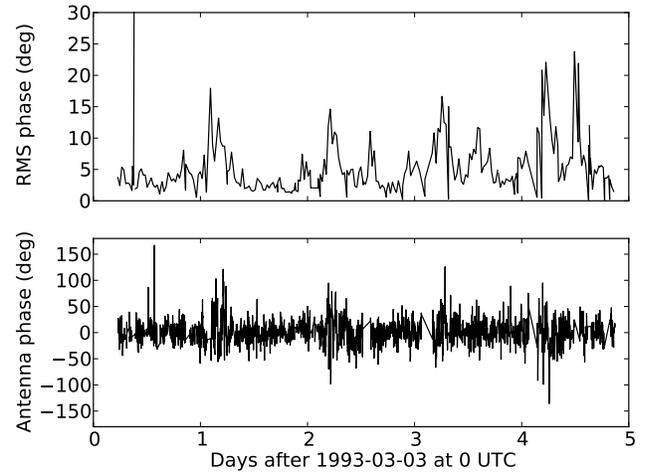}{\columnwidth}
\caption{The top panel shows the variation of $\sigma_{\phi}$ during the calibrator observations  at epoch 2. In the lower panel is shown the phase corrections for antenna 1, X-polarization determined from the self-calibration of target fields. Note the strong diurnal variation in both plots. Local noon is at 02$^h$ UT; the atmosphere at Narrabri is often at its most turbulent in the afternoon.}
\label{phasefluctuations}
\end{center}
\end{figure}

Secondary calibrators were observed for two minutes, and we characterise the phase fluctuations by calculating the standard deviation $\sigma_\phi$ of all measured phases during that interval.  Fig. \ref{phasefluctuations} shows the variation of $\sigma_\phi$ during the epoch 2 observations.  This pattern is also visible in the second-stage calibration results; the same figure shows phase corrections for antenna 1. 

\subsubsection{Field complexity}
These observations were designed with the assumption that most fields can be adequately described with very sparse measurement on the $uv$-plane.  The assumption is that most fields are empty apart from a small number of simply shaped sources.  At the wavelengths observed, the strongest source expected in an area of sky equal to the ATCA primary beam is 2.3~mJy \citep{Bridle:1989}, much less than the lower flux density limit of the PMN field selection.  However, a fraction of fields do not satisfy the assumption, and the ability to characterise the field suffers accordingly.  Two outstanding examples of such fields are PMN J1325$-$4257 (Centaurus A) and PMN J0519$-$4546 (Pictor A), two large, strong, and complex radio galaxies.

\subsubsection{A quantitative measure of data quality}
\begin{figure}
\begin{center}
\putfig{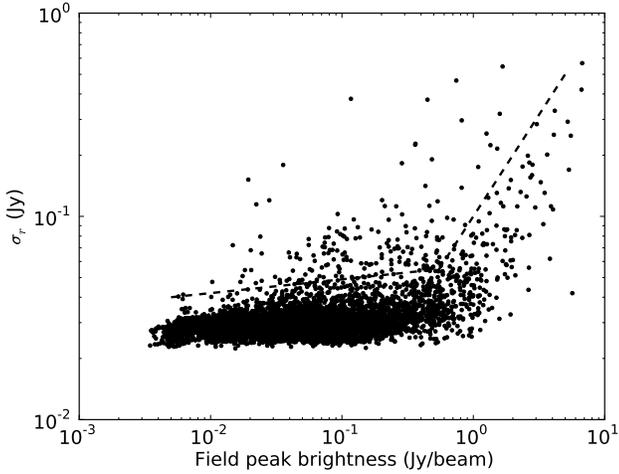}{\columnwidth}
\caption{The amplitude of residual noise $\sigma_r$ after the fitted source components are subtracted from the data.  The plot shows the dependence of $\sigma_r$ on the flux in the field, characterised here by the peak brightness $B_{max}$ in the image.  Note that the image noise corresponding to the quantity plotted here is $\sigma_{im} = \sigma_r/\sqrt{N_{vis}}$.  For most fields the number of visibility measurements $N_{vis} = 45$.   The dashed line indicates the threshold above which catalogue entries are given a data quality warning.} 
\label{residualnoise}
\end{center}
\end{figure}

In section \ref{modelfitting} above we describe the process of determining source characteristics from the data.  A product of this is a set of residual visibilities -- the original calibrated data less the contribution of the fitted components.  In the ideal case, these residual data would be the featureless noise that would correspond to an observation of an empty field; the amplitude of the noise would reflect the system noise with contributions from the receiver, losses in the telescope, ground radiation and diffuse sky emission.  In practice, the residuals are noisier and may have the signature of unfitted sources (perhaps out of the field).  Sources of additional noise are phase errors (instrumental or atmospheric),  confusion and contributions from sources whose brightness distribution is not well modelled as gaussian.  The noise increment in the residuals will increase with increasing source flux.  Fig. \ref{residualnoise} shows the $rms$ of the residual noise $\sigma_r$ and its dependence on the maximum brightness $B_{max}$ in the image of each field.  Note the sharp lower cutoff corresponding to the system noise. We indicate on that figure a threshold value for $\sigma_r$ above which entries in the final catalogue are flagged for lower reliability.

\subsubsection{Positional accuracy}
The positions of sources detected in each target field are determined with reference to the position of the secondary calibrator.  Thus there are two sources of error in the positions we determine for each source: the uncertainty in the position adopted for  the secondary calibrator, and the uncertainty in determination of the relative positions of target and calibrator.  The sources in the ATCA calibrator catalogue are categorised according to position uncertainty indicated in Table \ref{t_calquality}.  The present observations used 114 different secondary calibrator sources, 88 from class A, 3 from class B and 23 from class C.

\begin{table}
\begin{center}
\caption{ATCA calibrator source quality}
\label{t_calquality}
\begin{tabular}{ccl} \hline
Class & Accuracy & Measurement \\ \hline
A & $\sigma < 0\farcs02$ & VLBI positions \\
B & $0\farcs02 < \sigma < 0\farcs1$ & VLBI positions of resolved sources \\
C & $0\farcs1 < \sigma < 0\farcs25$  & ATCA measurement \\
\hline
\end{tabular}
\end{center}
\end{table}

\begin{figure}
\begin{center}
\putfig{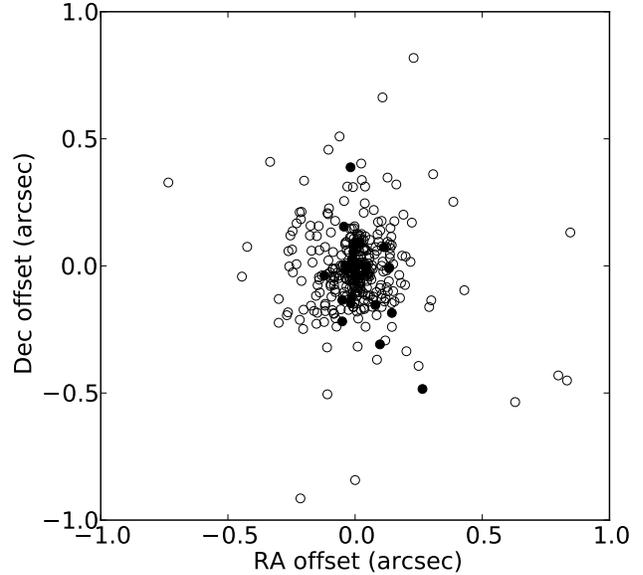}{\columnwidth}
\caption{Offsets of ATPMN positions from the 309 LCS1 (open) and 26 ICRF (filled) calibrators in our catalogue. The standard deviations of the errors in right ascension and declination for the LCS1 comparison are $\sigma_{\alpha} = 0\farcs16$, and $\sigma_{\delta} = 0\farcs17$. For the ICRF sources, these quantities are $\sigma_{\alpha} = 0\farcs08$, and $\sigma_{\delta} = 0\farcs15$.}
\label{astrerrs}
\end{center}
\end{figure}

The target fields included a number of sources that appear in catalogues of astrometric radio sources whose positions have been measured with an accuracy of a few milliarcseconds.  Both  the International Celestial Reference Frame (ICRF) \cite{Ma:1998cs} and the LCS1 catalogue of southern sources \cite{Petrov:2011fe} provide a good reference for this work, and the comparisons yield  good estimates of the position accuracy in this catalogue.  Fig. \ref{astrerrs} shows this comparison for 309 LCS1 sources and 26 ICRF sources for which our corresponding field used a class A secondary calibrator.  The observed position differences should be dominated by errors in transferring position from calibrator to target.  The standard deviations of the errors in right ascension and declination for the LCS1 comparison are $\sigma_{\alpha} = 0\farcs16$, and $\sigma_{\delta} = 0\farcs17$. For the ICRF sources, these quantities are $\sigma_{\alpha} = 0\farcs08$, and $\sigma_{\delta} = 0\farcs15$.

The errors in position determination depend on a number of factors, principally source flux density, confusion, source extent and phase stability.  The comparison with the two reference catalogues illustrates the source flux density dependence: the 25 ICRF sources in the comparison have minimum and mean flux densities of 0.32 Jy and 1.1 Jy; the LCS1 sources are much fainter and have minimum and mean flux densities of  0.04 Jy and 0.34 Jy.  The uncertainties in position reported by the {\sc uvfit}  process reflect all the factors listed above, being larger for sources that are weaker, extended, or lie in a field with other sources.  However, the magnitudes of the reported errors are considerably smaller than the spread observed in the comparisons with LCS1 and ICRF.  For each source in the catalogue we estimate the (1$\sigma$) uncertainties in RA and Dec, derived empirically from the values reported by  {\sc uvfit} the comparison with the astrometric references.  The median uncertainty is about 0\farcs4 in both coordinates.

\subsubsection{Flux density scale}
\label{radiometry}
\begin{figure}
\begin{center}
\putfig{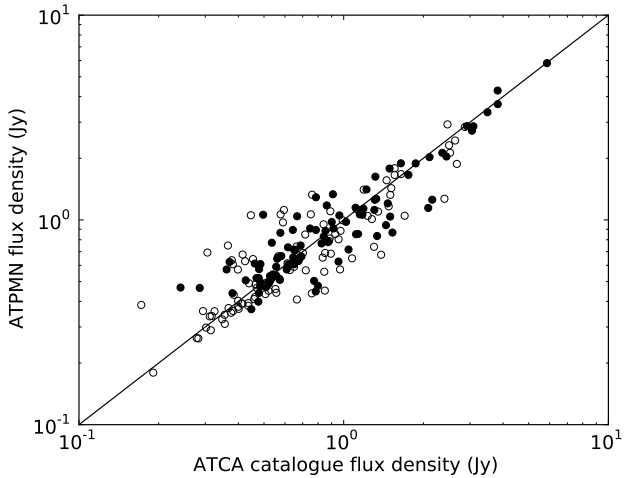}{\columnwidth}
\caption{Comparison of flux densities with ATCAT at 4.8~GHz (filled symbols) and 8.6~GHz (open symbols). The line shows equality; it is not a line of best fit.}
\label{atcalflux}
\end{center}
\end{figure}

All source flux densities are estimated relative to the flux density of the primary calibrator B1934$-$638.  We used the values of  \cite{Reynolds:1994vd}: $S_{4.8} = 5.829$ Jy and $S_{8.6} = 2.842$ Jy.   The flux density scale was transferred from the primary to secondary calibrator, and then from secondary to the target sources; each step can introduce errors in the final flux density estimate.  Because the secondary calibrators are strong point sources, they can be calibrated in a way that is relatively unaffected by phase fluctuations induced in the atmosphere (but see comments regarding this in \textsection\ref{calibration}).  The secondary calibrators were drawn from the ATCA calibrator catalogue (ATCAT);  many are known to be variable so a comparison of individual flux density measurements with ATCAT is not a valid check of our data quality.  Fig. \ref{atcalflux} shows that over the whole set of calibrators, our flux density measurements, at both observing frequencies, are consistent with ATCAT.

\begin{figure}
\begin{center}
\putfig{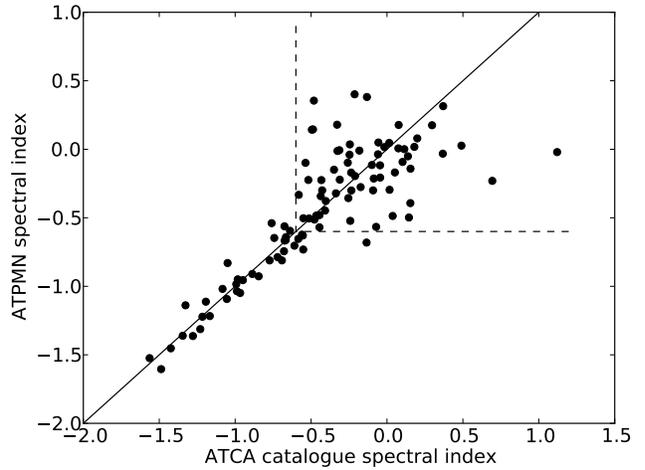}{\columnwidth}
\caption{Comparison of spectral indices $\alpha$ from 4.8 to 8.64 GHz  between ATPMN and ATCAT.  It is clear that sources with  $\alpha \ga -0.6$ have variable spectral indices.}
\label{atcalSI}
\end{center}
\end{figure}

Comparison of spectral index\footnote{We define the spectral index $\alpha$ through the expression for flux density as a function of frequency:  $S(f) = S(f_0) (f/f_0)^\alpha$.} ($\alpha$) provides a check on the relative consistency of the 4.8~GHz and 8.6~GHz calibration.  Fig. \ref{atcalSI} shows the results of that comparison: ATPMN spectral indices agree well with values from ATCAT where the spectral index is steep  $\alpha \la -0.6$.  The scatter for flat-spectrum sources is caused by variability, as discussed in \textsection \ref{variability}.

\begin{figure}
\begin{center}
\putfig{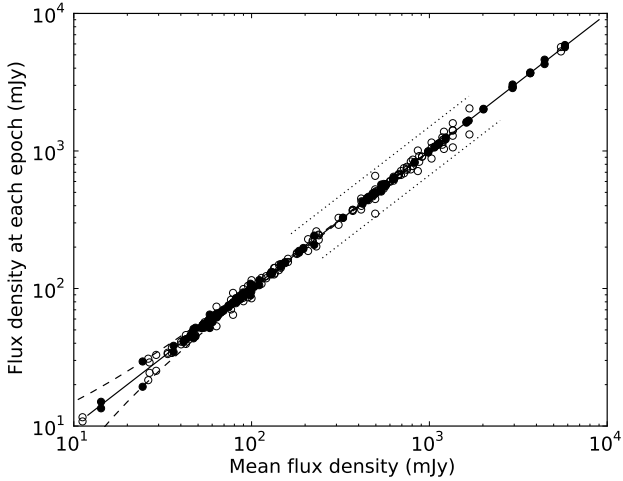}{\columnwidth}
\caption{Multi-epoch flux density comparisons at 4.8~GHz. Flux density measurements are plotted against the mean flux density for each  of 184 sources, filled symbols for sources with steep spectra ($\alpha^{8.6}_{4.8} < -0.5$) and open symbols for flat spectra.  The solid line shows equality, the dashed lines show deviations of a constant 10mJy and the pair of dotted lines mark two-fold flux density variations.}
\label{multiEpochFluxes}
\end{center}
\end{figure}

Another measure of the internal consistency of the ATPMN flux densities comes from the fields with repeated observations. 432 fields were observed at two or more epochs; 58 of these were secondary calibrators.  Many were re-observed because a break in the observing schedule caused one or more of their three cuts to be missed. In total 378 fields have  data from more than one epoch.  We selected from these 158 sources that are the brightest in their field, are compact (size $<$ 1\arcsec) and are detected at both observing bands.  Analysis of these data show a high degree of consistency in flux density scale over the five observing sessions. Fig. \ref{multiEpochFluxes} shows the individual flux density measurements against the mean flux density of each of these sources. In the plot we distinguish between sources with steep spectra and and those with flat spectra.  The data are consistent with source variability of up to a factor of two (more pronounced amongst flat-spectrum sources), a random measurement error of a few milli-Janskys, and small systematic changes in the epoch-to-epoch flux density scale.  Assuming the flux density of steep spectrum sources to be constant, we approximate  the error in flux density measurement as $\sigma_{S} = \Delta S + \epsilon S$, where $(\Delta S, \epsilon)$ are  (6.6 mJy, 1.6\%) and  (10 mJy, 4.5\%) at 4.8~GHz and 8.6~GHz respectively.

Amongst the target fields with multiple observations was the source PMN~J1939$-$6342 (B1934$-$638), the source used for primary flux density calibration of the ATPMN data.  Data from the epoch 1 and 2 observations of J1939$-$6342 were processed in the same way as all other targets, with the flux density scale being transferred from B1934$-$638 to secondary calibrators (epoch 1: J1549$-$790, J2146$-$783, J1624$-$61; epoch 2: J2030$-$689), and then on to the target data.  The results for  J1939$-$6342 (Table \ref{flux1939} are consistent with the the results given above for multi-epoch fields.

\begin{table}
 \centering
 \caption{Measured flux densities of PMNJ1939$-$6342, relative to the values assumed for B1934$-$638:   $S_{4.8} = 5.829$ Jy and $S_{8.6} = 2.842$ Jy}
 \label{flux1939}
\begin{tabular}{ccccc}
  \hline
   Epoch    & \multicolumn{2}{c|}{4.8~GHz} & \multicolumn{2}{c|}{8.6~GHz} \\
                &   (mJy) & $\Delta$   & (mJy) & $\Delta$           \\
\hline
1 & 5893 &  1.1\% & 2888 &  1.7\% \\
2 & 5675 & -2.7\% & 2758 & -2.9\% \\
\hline
\end{tabular}
\end{table}

\begin{figure}
\begin{center}
\putfig{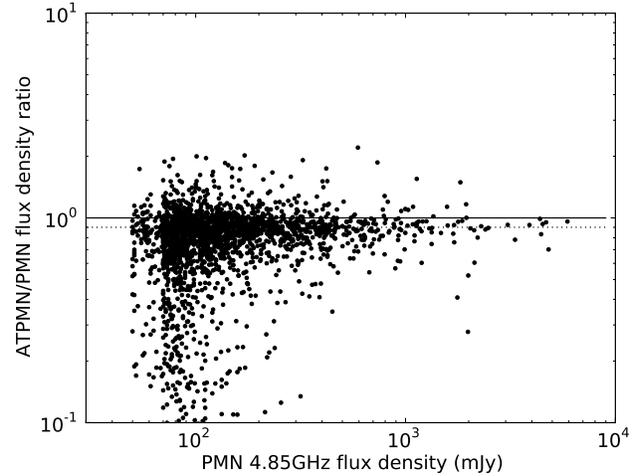}{\columnwidth}
\caption{The ratios of flux densities measured  in the ATPMN (4.8~GHz) and  PMN (4.85~GHz) catalogues.  The selection is of sources unresolved with the ATCA and appearing as the only source in the field.  The solid line marks flux density equality, and the dotted line shows a 10\% depression of ATCA flux densities.  The lower density of points below PMN flux density of 70~mJy results from the lower flux density threshold for ATCA follow-up selection south of $\delta = -73\degr$.  Points with very low flux density ratios correspond to fields with extended sources visible in the PMN data and a faint coincident point source detected by the Compact Array.}
\label{PMN_pointFluxes}
\end{center}
\end{figure}

We make two further comparisons of the flux densities measured for the ATPMN sources, first at 4.8~GHz with the PMN catalogue, and then at 8.6~GHz with VLA measurements reported in the CRATES catalogue \citep{Healey:2007fa}.  Fig. \ref{PMN_pointFluxes} shows the ratios of ATPMN and PMN flux densities for the sample of fields containing only one unresolved ATPMN source.  There is a wide scatter, presumably because of source variability and the insensitivity of the ATCA to extended emission. In spite of the scatter a flux density scale difference is evident, Parkes flux densities being consistently greater than those measured with the ATCA.  The PMN flux density scale \citep{Griffith:1993cl} was determined from a comparison of a set of sources listed by \cite{Kuehr:1981wo} 1981, thereby tying it to the flux density scale of \cite{Baars:1977va} 1977.  This discrepancy is consistent with an additive error to PMN source flux-densities of between 5~mJy and 10~mJy and a flux scale difference between the Parkes and the ATCA measurements of approximately 6 per cent.

\begin{figure}
\begin{center}
\putfig{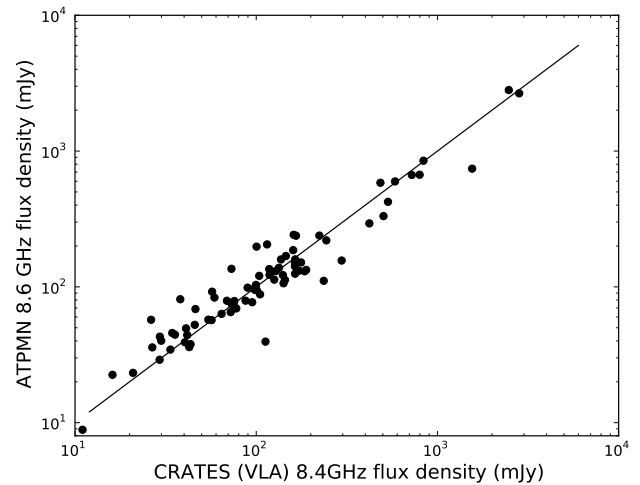}{\columnwidth}
\caption{Comparison of source flux densities in the CRATES (VLA at 8.4~GHz) and ATPMN (at 8.6~GHz) catalogues}
\label{CRATESfluxes}
\end{center}
\end{figure}

The CRATES catalogue is an amalgamation of radio source information form a number of single-dish and interferometer surveys.  We have looked at sources in the declination band common to the ATPMN survey and VLA visibility $-41\degr < \delta < -38\degr$.  In that region there are 79 sources with flux density measured at both the VLA (8.4GHz) and the ATCA (8.6~GHz).   Fig. \ref{CRATESfluxes} shows the comparison. The flux-weighted mean and median of the ratios of ATCA to VLA flux density are 1.02 and 1.01 respectively.

\subsubsection{Polarimetry}
\label{polarimetry}

All observations were made using the full polarimetric capabilities of the ATCA.  A complete analysis of the polarimetric properties of the ATPMN sources will be the subject of further work, and the results made available in a future version of the catalogue.

\subsubsection{Resolution and angular size discrimination}
\label{resolution}
\begin{figure}
\begin{center}
\putfig{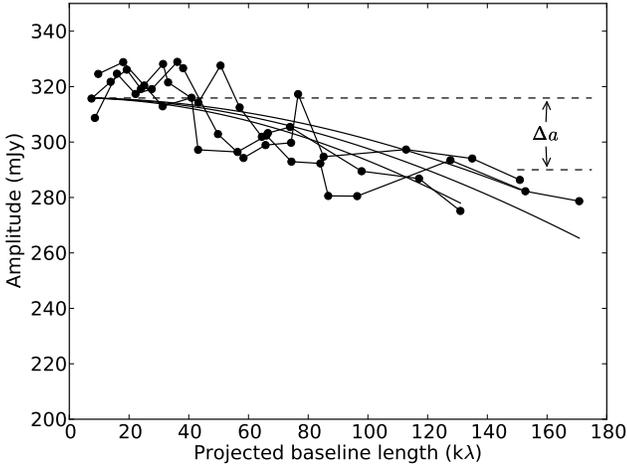}{\columnwidth}
\caption{Visibility amplitudes for a small diameter but resolved source,  J005645.8$-$445102  (PMN J0056$-$4451).  Data at 8.6~GHz from all three observations are shown, with the modelled visibilities shown by the continuous curved lines.  The best-fit angular size is $0\farcs5 \times 0\farcs2$ with a position angle of $30\degr$.  We regard a source resolved if the depression of long baseline amplitudes, shown here as the vertical distance between the dashed lines, exceeds $\sigma_{vis}/\sqrt{N_{vis}}$.}
\label{resolutionlimits}
\end{center}
\end{figure}

The typical synthesised beam sizes for these data are 2.2~$\times$~1.6~arcseconds and 1.3~$\times$~0.9~arcseconds at the two observing frequencies.  However, it is possible to measure sizes of sources with smaller angular diameters than the synthesized beam. The visibility function of small but non-zero diameter sources differs measurably from that of a true point (unresolved) source.  The size of the smallest, noticeably resolved source depends on the strength of the source and the distribution of baseline-lengths used in the measurement.  Our criterion for whether or not a source is resolved is illustrated in Fig. \ref{resolutionlimits}.  The expected visibility amplitude on a baseline of length $D$ for a gaussian source of diameter $W$ is reduced from its zero-baseline value by $\Delta a$:
\[
\Delta a = S \exp [-(D W)^{2}]
\]
We do not claim a source to be resolved unless $\Delta a > \sigma_{vis} \ \sqrt{N_{vis}}$, where $\sigma_{vis}$ is the uncertainty in each visibility measurement and $N_{vis}$ is the number of visibility measurements for the source.  For most fields, $\sigma_{vis} \sim 30$mJy and $N_{vis} = 45$.

The angular size distribution of sources in our sample is strongly influenced by the instrumental characteristics of the ATCA.  At 4.8~GHz the synthesized beam is typically  $2\farcs0 \times 1\farcs5$.  Sources with angular size smaller than some flux-dependent fraction of the synthesized beam cannot be distinguished from points, as discussed above.  

\begin{figure}
\begin{center}
\putfig{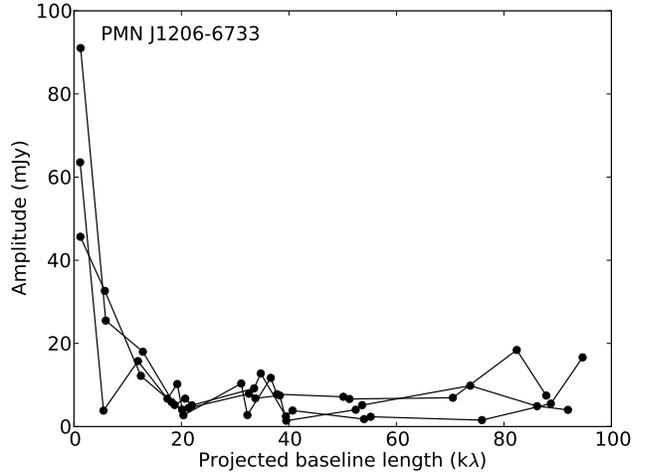}{\columnwidth}
\caption{Visibility amplitudes for the field containing the 95~mJy source PMN J1206$-$6733.  Data at 4.8~GHz from all three observations are shown.  This source, which lies at Galactic latitude $b = -5.1\degr$,  appears in the PMN catalogue as unresolved; the ATCA visibility measurements are consistent with a source whose extent exceeds 20\arcsec.}
\label{visPlot0}
\end{center}
\end{figure}

\begin{figure}
\begin{center}
\putfig{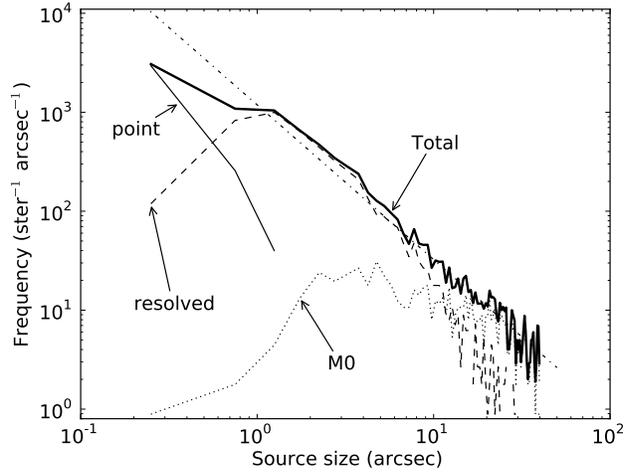}{\columnwidth}
\caption{The differential distribution of angular sizes for sources in the sample, as measured at 4.8~GHz and normalised by the solid area of the sky surveyed.  The labelled lines show point sources (solid), resolved sources (dashed), too extended for imaging (dotted) and the total distribution (heavy solid).  }
\label{angularSizes}
\end{center}
\end{figure}

On the other hand, sources with large angular sizes have measureable visibility amplitudes on none, or only the shortest of the ATCA baselines.  All fields observed correspond to PMN sources for which the Parkes telescope measured at least 50mJy of flux density; non-detection with the all ATCA baselines means the source size is comparable or larger than the Array's largest fringe $\Delta\theta_{max}$.  At 4.8~GHz,  $\Delta\theta_{max} \simeq $ 0\farcm6, 1\farcm3  and 2\farcm2  for the 6D, 6C and 6A configurations used for these observations (see Table \ref{sessions}).  If flux is detected on the shortest baseline, it can be used with the PMN flux density measurement to estimate the angular size of the source.    About 60 per cent of fields with no imaged source do have significant flux on the shortest baseline; Fig. \ref{visPlot0} shows the visibility amplitudes for such a field.  Fig. \ref{angularSizes} shows the angular size distribution of these sources of intermediate extent, and of the successfully modelled resolved and unresolved sources.

\section{The Data Catalogue}
\label{theCatalogue}
We present the main result of our work as a catalogue of sources. All identified components are listed separately; there has been no attempt to associate components to physical objects, although each record includes the name of the PMN source observed.  In many cases, multiple sources observed in a single PMN field are components of a single object; in some cases they result from chance alignment.

In the formation of the final source list, we have excluded some sources resulting from the data analysis.  We have omitted all sources with peak brightness at 4.8~GHz of less than 6~mJy/beam (close to five times the image {\em rms} for most fields.  We have also excluded all sources further than 5 arcminutes from the field centre, at which point the sensitivity of the 8.6~GHz primary beam of the ATCA is one tenth of its peak.

The ATPMN catalogue gives the position, flux densities at 4.8~GHz and 8.6~GHz, source size parameters, spectral index and the epoch of observations for each source.  In all cases, source parameters are derived from the model-fitting to the visibility data as described in \textsection\ref{imagingandfitting}, not from an analysis of the radio images.  Table 5 shows a sample of 30 catalogue entries ; the full catalogue is available on line on Vizier.

\addtocounter{table}{1}

The columns are:
\begin{description}
\item[(1)] ATPMN source name
\item[(2)] PMN source name of parent field
\item[(3-4)] Right Ascension and Declination (J2000). The accuracy is indicated by the number of digits given: in each case the error is less than 10 in the final digit.
\item[(5-6)] Flux density at 4.8~GHz (mJy), and its uncertainty.
\item[(7-8)] Flux density at 8.6~GHz (mJy), and its uncertainty.
\item[(9-11)] Source size at 4.8~GHz as major and minor axes (arcseconds) and position angle (degrees)
\item[(12-14)] Source size at 8.6~GHz as major and minor axes (arcseconds) and position angle (degrees)
\item[(15-16)] Spectral index $\alpha_{8.6}^{4.8}$ and its uncertainty.
\item[(17)] Epoch of observation (see Table \ref{sessions}).
\item[(18,19)] Quality flags as described in \textsection\ref{dataqualityflags}.

\end{description}

The spectral index and its uncertainty are calculated as:
\[
\alpha = \frac{\log (S_6/S_3)}{\log (f_6/f_3)},
\epsilon_{\alpha} = \frac{\sqrt{(\epsilon_{S_6}/S_6)^2 + (\epsilon_{S_3}/S_3)^2}}{\log (f_6/f_3)}
\]
where the subscripts denote the observing bands at 6~cm (4.8~GHz) and 3~cm (8.6~GHz), $S$ is the measured flux density, and $f_6 = 4.8$~GHz, $f_3 = 8.6$~GHz.

\subsection{Data quality flags}
\label{dataqualityflags}
The reliability of tabulated source parameters is affected by many factors as discussed in \textsection\ref{dataquality}.   We flag catalogue entries with less reliability according to just two criteria: model-fitting residual and the number of visibility points used in the fit.   To each criterion corresponds a flag digit that is normally `0'.  Fig. \ref{residualnoise} shows the fit residuals for each field and how they depend on the peak brightness. All sources from fields with points above the threshold indicated in the figure are flagged with figure `1' in the first flag.  Normally, fields were observed with three scans at different hour angles to give 45 visibility measurements. A few fields (166) has less than 20 visibility measurements; these were not included in the processing. Fields with 30 or fewer points have measurements at only two hour angles;  the catalogue entries corresponding to these are flagged with the figure `1' in the second flag.

\section{Source population}
\label{sourcepopulation}
It has been customary to use radio sky surveys to compute differential source counts, the number of sources visible within a flux density range per unit of sky area.  Historically, the analysis of the differential source count over a range of radio frequencies and flux densities has been used to rule out several possible models for the shape of the universe and has led to the belief that there has been a marked change in the population of objects responsible for the visible radio emission.  In general the uncertainties in differential source counts do not adhere to the values expected from pure statistical considerations.  Rather they are dominated by the dependence of source counts on calibration, and beam and resolution corrections (ultimately the angular spectrum of brightness sensitivity) \citep{DeZotti:2010gj}.  The comparison of the ATPMN and PMN catalogues provides a demonstration of this difficulty, and such a comparison may contribute to its resolution for the case of 4.8~GHz source counts. 

 We begin our overview (\textsection \ref{fieldmorphology}) of the source population by summarising their basic parameters and make a simple numerical comparison: how many ATPMN sources correspond to each PMN source.  In \textsection \ref{othercatalogues} we look at the degree of overlap between ATPMN sources and those in other source catalogues.  In \textsection \ref{variability} we examine the variability of source flux density and spectral index.

\subsection{Field morphology}
\label{fieldmorphology}

\begin{table}
 \centering
 \caption{The distribution of field among the morphology types [M0, M1, ... M6 ] as defined  in the text.  The 166 fields with insufficient data for full analysis are excluded, reducing the total number in this table to 8219.}
\begin{tabular}{clll}
  \hline
   & Morphology    & $N_{4.8}$  & $N_{8.6}$ \\
\hline
0 & no identifiable source & 1441 & 2138 \\
1 & single point source & 2730 & 2745 \\
2 & single resolved source & 2162 & 2009 \\
3 & two point sources & 158 & 199 \\
4 & one point, one resolved & 425 & 424 \\
5 & two resolved sources & 1056 & 603 \\
6 & more than two sources & 247 & 101 \\
\end{tabular}
  \label{morphology}
\end{table}

\begin{figure}
\begin{center}
\putfig{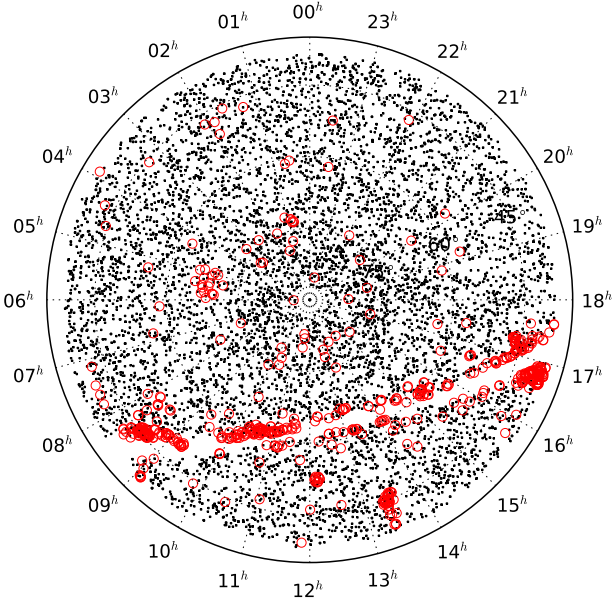}{\columnwidth}
\caption{The distribution of ATPMN sources (black points) on the sky, shown in equatorial coordinates centred on the south celestial pole.  The open red symbols mark the positions of fields for which no significant flux density was measured on the shortest ATCA baseline, an indication that the corresponding PMN source is extended (see \textsection \ref{fieldmorphology}).  Most of these lie close to the Galactic Plane, one of the Magellanic Clouds or Centaurus-A.}
\label{skyPlot}
\end{center}
\end{figure}

From the ATCA observations of  \Nfields PMN source positions we have detected \Nsources individual components.  Their distribution on the southern sky is shown in Fig. \ref{skyPlot}.  We have characterised each field using a simple classification scheme defined in Table \ref{morphology}  which also reports the numbers of fields in each morphology classes M0$-$M6.  Fields of class M0 themselves fall into two categories: those with some detectable flux density on the shortest ATCA baselines, and those without as discussed in \textsection\ref{resolution}.    Fig. \ref{skyPlot} shows the distribution of fields with no detectable flux on the ATCA, lying primarily along the Galactic Plane, in the Magellanic Clouds and close to the near-by radio-galaxy Centaurus-A (NGC~5128).  The processing of the original PMN survey involved high-pass filtering of the scan data to better identify compact sources \citep{Condon:1993kf}.  Evidently, this filtering was less effective in regions of strong extended emission and has resulted in some entries in the PMN catalogue corresponding to data processing artefacts. Of the 589 fields with no detectable flux, 407 (69 per cent) lie within 5 degrees of the Galactic Plane.

\begin{figure}
\begin{center}
\putfig{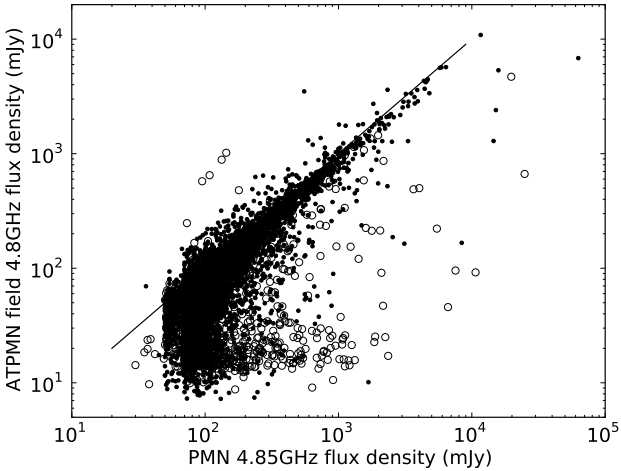}{\columnwidth}
\caption{Total ATCA flux detected at 4.8~GHz against the 4.85~GHz PMN flux density for all PMN fields observed.  The open symbols show fields with no source identified in the image; the ATPMN flux density in those cases is the maximum visibility amplitude observed.}
\label{PMN_fluxes}
\end{center}
\end{figure}

Fig. \ref{PMN_fluxes} gives another view of the comparison between ATCA and Parkes views of the PMN sources, showing the total flux detected with the ATCA against the PMN flux for each field.  The effect of missing extended flux in the ATCA observations is clear in the tendency of points to fall below the equality line.  The excess of points close to the ATCA noise level ($\sim$30~mJy) results from the PMN sources with no detectable ATCA flux and are likely to be the data processing artefacts in the PMN catalogue mentioned above.

\begin{figure*}
\begin{center}
\putfig{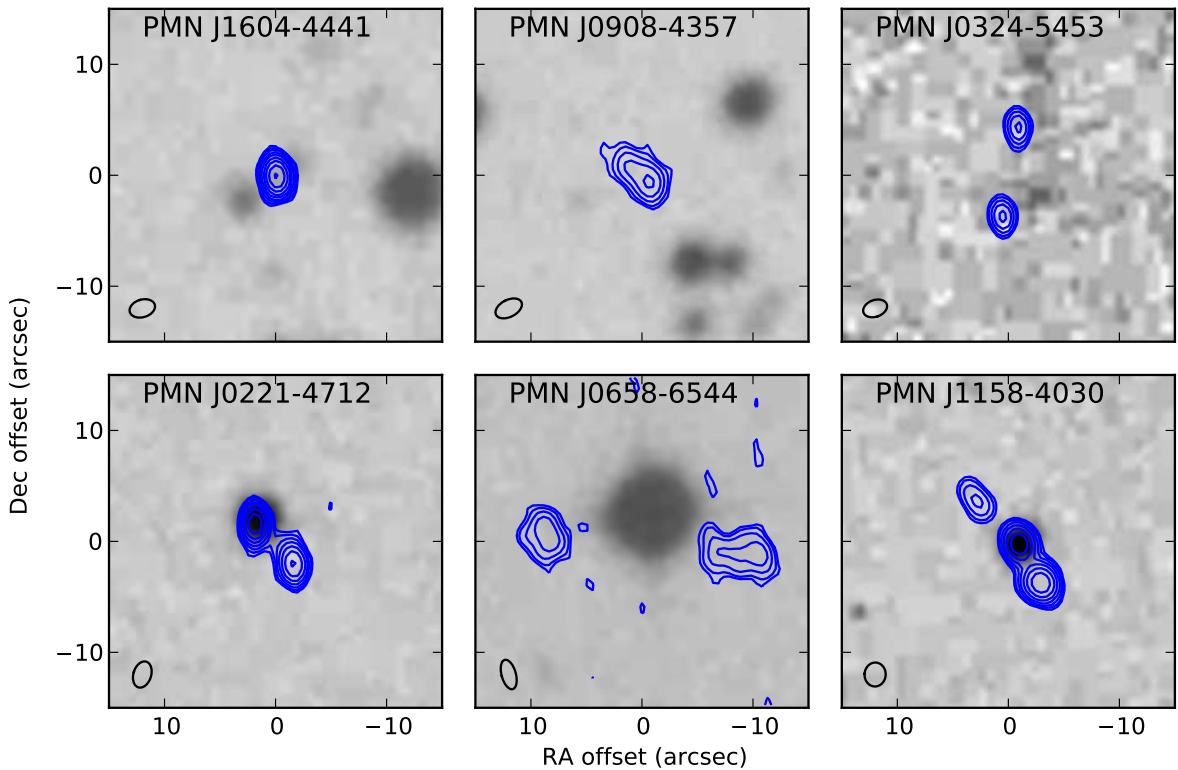}{\textwidth}
\caption{Example fields chosen from each morphology type M1 -- M6; source properties for each field are given in Table \ref{sourceProperties}. Each image is a 30 arcsecond square and centred on the mean source position for the field (not the PMN source position);  contours of the 4.8~GHz ATCA image are overlaid on the corresponding image from the UK Schmidt Blue (IIIaJ) survey, downloaded from the SuperCOSMOS Sky Surveys archive. The ATCA synthesized beam at 4.8~GHz is indicated at lower left in each image. }
\label{big_examplefields}
\end{center}
\end{figure*}

\begin{figure}
\begin{center}
\putfig{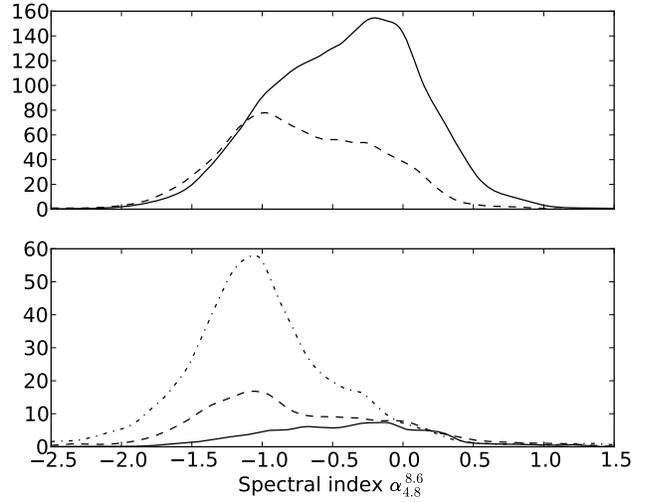}{\columnwidth}
\caption{Spectral index distributions for single-source fields (top) and sources in double-source fields (bottom). Top: morphology types 1 (point sources -- solid) and 2 (resolved sources -- dashed). Bottom: morphology types 3 (two point sources -- solid), 4 (one point, one resolved source -- dashed) and 5 (two resolved sources -- dash-dotted).  The distributions were constructed as the sum of gaussian functions for each source; each is centred on the spectral index for its source and has half-width equal to the 1-$\sigma$ uncertainty, and has unit area.}
\label{si_by_morph}
\end{center}
\end{figure}

In Fig. \ref{big_examplefields} we display images from an example of each morphology type (M1 - M6); the parameters of each source appear in Table \ref{sourceProperties}.  To summarise the characteristics of the different morphologies, we show their distributions of angular size, spectral index, and for the double sources, angular separation.  In Fig. \ref{si_by_morph} we show the distribution of spectral indices for sources in fields of the various morphologies.  These distribution plots have been formed from sources for which the uncertainty in spectral index was $\epsilon_{\alpha} < 0.5$.  As expected, there is a marked tendency for the unresolved sources (morphology types 1 and 3) to be rich in flat spectra ($\alpha > -0.5$), and for the extended sources to have steep spectra.

\begin{table}
 \centering
 \caption{Properties of the sources appearing in Fig. \ref{big_examplefields}}
  \label{sourceProperties}

\begin{tabular}{clcrr}
  \hline
  \multicolumn{1}{c}{PMN}  & Angular & $S_{4.8}$   & $\alpha^{8.6}_{4.8} $\\
  \multicolumn{1}{c}{name}          & size  & (mJy) & \\
\hline
PMN J1604-4441 & point & 1216 & +0.3 \\
PMN J0908-4357 & $3\farcs7 \times 1\farcs2$ & 251 & $-$1.0 \\
PMN J0324-5453 & point & 137 & $-$0.9 \\
                            &  point & 180   & $-$1.3   \\
PMN J0221-4712 & point & 117 & $-$1.0 \\
                            &  $1\farcs3 \times 0\farcs6$ & 135   &   $-$0.5 \\
PMN J0698-6544 &  $5\farcs6 \times 2\farcs3$ & 307 & $-$1.2 \\
                            &  $3\farcs3 \times 2\farcs9$  & 234  & $-$1.3   \\
PMN J1158-4030 & $1\farcs1 \times 0\farcs8$  & 165 & $-$2.2 \\
                            &   $1\farcs3 \times 0\farcs7$  & 100 &   $-$0.1    \\
                            &  $2\farcs1 \times 0\farcs2$  &   37 &    +0.2         \\
 \hline
\end{tabular}
\end{table}

\subsection{Comparison with other catalogues}
\label{othercatalogues}
Before looking at other surveys, we make some comparisons between the ATPMN catalogue and the unpublished \PMNCA\ catalogue prepared by N. Tasker  \citep{Tasker:1997} and described by \cite{Wright:1997wz}.  The \PMNCA\ catalogue comprises 7178 sources, compared with the \Nsources in ATPMN.  Of those sources, 6183 lie within 80 arcseconds of an ATPMN position; the difference, 970 sources do not appear to have an ATPMN counterpart within that radius. Of those 6183 pairs, 316 are separated by more than 3 arcseconds, and 189 by more than 10 arcseconds.  This is expected, given the calibrator position errors we found in the data (\textsection\ref{calibration}).  The mean position difference between sources in the two catalogues is 0\farcs27 in Right Ascension, and 0\farcs10 in Declination. In a flux density comparison we found a flux density scale difference of about 10 percent, the \PMNCA\ flux densities being greater than ATPMN flux densities. The flux scale difference is consistent with the different values used for the flux density of PKS~B1934-638 in the two analyses: $S_{4.8} = 6.33$~Jy by \citeauthor{Wright:1997wz} for \PMNCA\ and  $S_{4.8} = 5.83$~Jy in this work.   We also found a large scatter in flux densities, particularly for sources weaker than about 200~mJy.

We have searched several other source catalogues for matches to ATPMN sources.  Here we summarise these comparisons for the Sydney University Molonglo Sky Survey (SUMSS, \citealt{Mauch:2003ed}) and the 2nd epoch Molonglo Galactic Plane Survey (MGPS-2, \citealt{Murphy:2007kx}), the Australia Telescope 20GHz survey (AT20G, \citealt{Murphy:2010ej}) and the Second Fermi Large Area Telescope catalogue of Gamma-ray sources (2FGL, \citealt{Collaboration:2011tta}).  Future work will include searches for ATPMN counterparts in optical images.

\subsubsection{SUMSS and MGPS-2}
\begin{figure}
\begin{center}
\putfig{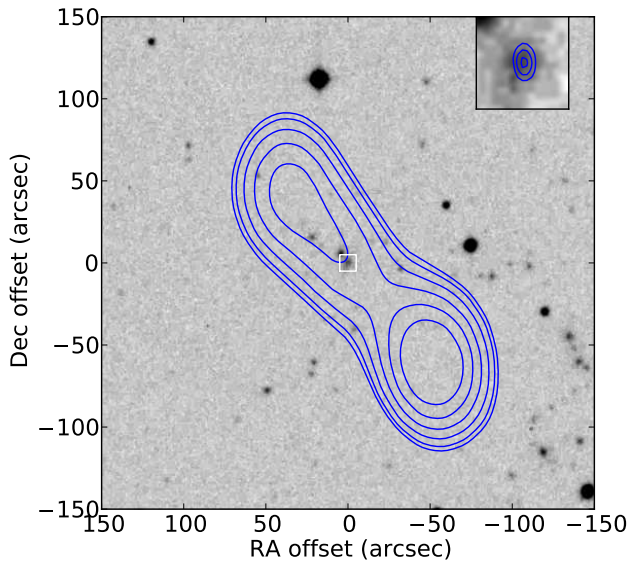}{\columnwidth}
\caption{Optical and radio images centred on the position of ATPMN J234237.7-443402 ($\alpha = $~23:42:37.7, $\delta = -$44:34:02.0).  The central square marks the position of the 10 arcsecond wide inset image.  In the main image, the SUMSS brightness contours at 50, 60, 80, 110, 150, 200  mJy/beam.  In the inset, ATPMN contours at 6,14,25 mJy/beam.  In both cases the contours are overlaid on the corresponding image from the UK Schmidt Blue (IIIaJ) survey, downloaded from the SuperCOSMOS Sky Surveys archive. The ATPMN source has flux density $S_{4.8} = 43$mJy and $S_{8.6} = 30$ mJy. }
\label{SUMSScomposite}
\end{center}
\end{figure}

The SUMSS and MGPS-2 catalogues were derived from surveys using the Molonglo Radio Telescope (MOST) at 843~MHz. They cover complementary parts of the sky, although the MGPS-2 catalogues only sources that are unresolved by the MOST; together they include all the sky observed for ATPMN.  The resolution of MOST is 45 arcseconds, and the uncertainty in position determinations in SUMSS for bright point sources is $\sim$2 arcsec \citep{Mauch:2003ed}.  We have associated ATPMN sources with sources in SUMSS and MGPS-2 if the positions are within 30 arcseconds.   Of the \Nsources sources listed in ATPMN, 6876 have SUMSS counterparts and 1693 correspond to MGPS-2 sources.  This leaves 472 ATPMN sources not found within the 30 arcsecond radius in either SUMSS or MGPS-2.  Further investigation shows that the number of plausible associations increases as the radius is increased until chance associations begin to dominate above a radius of 50 arcseconds. After the radius increase, about 250 ATPMN sources remain unassociated.  Inspection of the associations with large apparent position differences shows that many arise from the ATPMN lying at the core of a radio galaxy and the SUMSS sources, typically a pair, being the extended radio lobes.  Fig. \ref{SUMSScomposite} shows an example of such a system.

\subsubsection{AT20G}
\begin{figure}
\begin{center}
\putfig{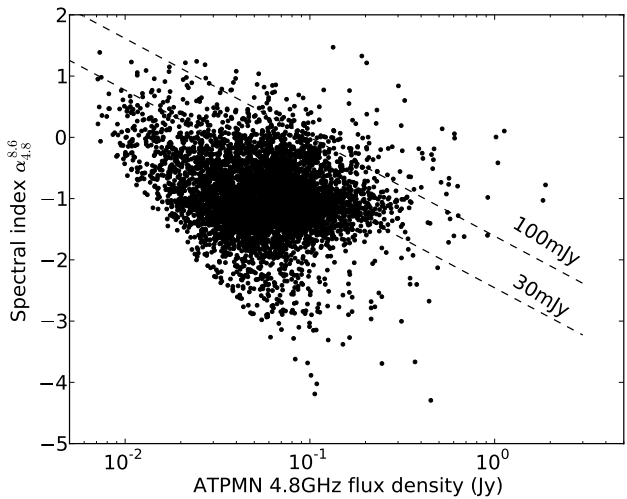}{\columnwidth}
\caption{7069 ATPMN sources do not lie within 5\arcsec of any AT20G source; for 5165 of these we have flux density estimates at two frequencies and so can infer their likely flux density density at 20GHz.  Here we show the ATPMN sources for which we have spectral indices in comparison to the inferred 20GHz flux density. Very few sources lie beyond the 100~mJy line, the level quoted as 91 per cent complete in AT20G.}
\label{nonAT20G}
\end{center}
\end{figure}

The AT20G catalogue is a list of sources selected from observations at 20~GHz; it is 91 per cent complete for sources above 100~mJy at 20~GHz.  Over the part of the sky common to  ATPMN, there are 2365 AT20G sources.  2093 of these lie within 5\arcsec of an ATPMN source.  Fig. \ref{nonAT20G} shows the flux densities and spectral indices of ATPMN sources that are not listed in AT20G.  The inferred 20~GHz flux density is indicated on the plot; the upper boundary of the points corresponds to the flux density limit of the AT20G survey.  

A fraction (422 out of 2365 -- 17\%) of AT20G sources in the ATPMN sky have no ATPMN counterpart within 5\arcsec.  A little more than half of these have no 5~GHz flux density in the AT20G follow-up or have a 5GHz flux density of less than 70~mJy, the ATPMN selection threshold.  The balance are either resolved as double by ATPMN, with each component falling outside the 5\arcsec radius, or lie in a PMN field but are not detected for some reason: very strong source in field, or extended source dominating so that our source-finding algorithm misses weak point sources in the field.  Because of the different resolution of the two surveys, the effects of source extent and source variability with time can become confused.  However, for cases in which the ATPMN flux density is greater than the AT20G measurement for the same source, variability is the most likely cause.  Approximately 20 per cent of flat-spectrum sources measured in both surveys have ATPMN 8.64~GHz flux density more than 1.5 times the corresponding AT20G value.  Assuming symmetry, 40 percent of flat-spectrum sources changed in 8.6~GHz flux by more than a factor of 1.5 over the 11-year interval between measurements.

\begin{figure}
\begin{center}
\putfig{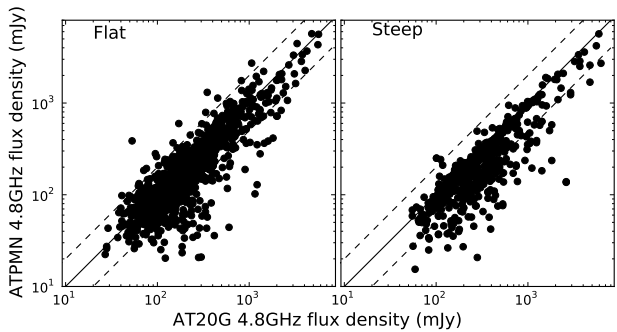}{\columnwidth}
\caption{Comparison of ATPMN and AT20G flux density measurements at 4.8~GHz. 1252 flat spectrum sources (left) and 713 sources with steep spectra (right) are shown.  The solid lines show equality and the dashed lines mark factors of two stronger and weaker.  The scatter amongst the flat-spectrum sources is partly attributed to variability; epochs of the two observations differed by 11 years.  Many of the steep-spectrum sources appear weaker in the ATPMN catalogue; we attribute this to the different spatial-frequency sensitivity of the two surveys: ATPMN used 6~km arrays whereas AT20G used the sorter 375~m arrays and was sensitive to more extended flux.}
\label{at20gFluxes}
\end{center}
\end{figure}

In Fig. \ref{at20gFluxes} we compare the ATPMN and AT20G flux densities measured at 4.8~GHz.  Two spectral classes are plotted: 1252 flat spectrum sources and 713 sources with steep spectra  are shown.  The ruled lines show flux density equality (solid) and factors of two different (dashed). It is clear that the relation is tighter for steep spectrum sources. Deviations from equality can be explained by measurement error,  different spatial-frequency sampling (AT20G was more sensitive to extended emission), or source variability which is most marked for flat spectrum sources.

\subsubsection{Fermi LAT}
\label{fermilat}
\begin{figure*}
\begin{center}
\putfig{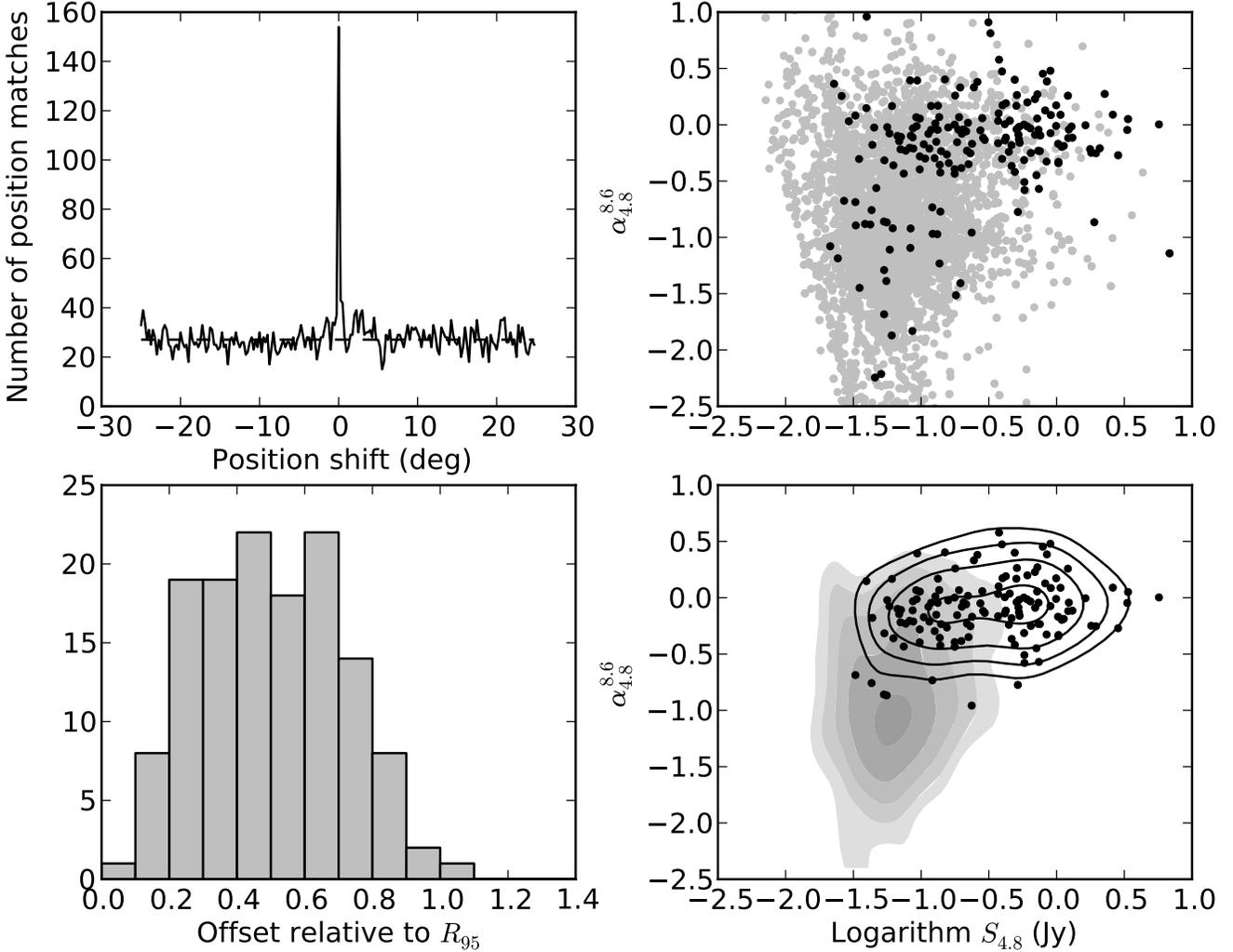}{\textwidth}
\caption{Analysis of the associations between the Fermi LAT 2FGL and ATPMN catalogues.  Upper left: Number of 2FGL positions  within $1.4 R_{95}$ of an ATPMN source, shown as a function of a RA position shift $\Delta\alpha$ of the ATPMN sources.  Upper right: ATPMN sources in the  $\Delta\alpha = 0$ matched set \F0 (black) and the  $\Delta\alpha \neq 0$ matched sets \Fda (grey), plotted on the ($S_{4.8GHz}, \alpha^{8.6}_{4.8}$) plane. Lower right: probability density functions $P_p$ (black contours) and $P_{\Delta\alpha}$ (grey scale); the selected likely candidates are shown as black points.  Lower left: the distribution of offsets in units of $R_{95}$ of the final selected matches.}
\label{fermiCorr}
\end{center}
\end{figure*}

The Second Fermi-LAT catalogue (2FGL) lists 1873 $\gamma$-ray sources identified from the initial 24 months of operation of the  Large Area Telescope (LAT) on the Fermi Gamma-ray Space Telescope \cite{Collaboration:2011tta}.  The 2FGL lists 330 sources in the ATPMN sky.  Amongst these we have sought those for which an ATPMN source lies within  $1.4 R_{95}$ (where $R_{95}$ is the 95 per cent error radius).  Because of the relatively large 2FGL error radius (typically $\sim 10$ minutes of arc), we expect a significant number of position matches between 2FGL and ATPMN sources that do not arise from physical association.  To gauge this coincidence rate, we have counted apparent associations for a series of trials in which the position $(\alpha,\delta)$ of each ATPMN source is shifted to $(\alpha + \Delta\alpha_{0}/\cos{\delta},\delta)$ for a range of $\Delta\alpha$ values.  The top left panel of Fig. \ref{fermiCorr} shows the number of  2FGL positions lying within  $1.4 R_{95}$  of at least one ATPMN source, as a function of $\Delta\alpha$.  Referring to the two sets of ATPMN-2FGL matches as \F0, the real matches at $\Delta \alpha = 0$ and \Fda the test matches with $\Delta\alpha \neq 0$, \F0 has 154 elements and \Fda has $27\pm5$ (normalised by the number of position shifts); this indicates a likely $127\pm5$ real radio\,$-$\,$\gamma$-ray source associations.  

The set \F0 holds radio--$\gamma$-ray associations and accidental alignments in the approximate ratio 127:28.  To form a list of the most likely associations we note that the radio sources in \F0 and \Fda  are differently distributed in both flux density and spectral index.  We estimate the probability density functions (PDF) of the source coordinates in $S_{4,8}$, $\alpha^{8.6}_{4.8}$ space; we denote these distributions as $P_0$ and $P_{\Delta\alpha}$ for the sets \F0 and \Fda  respectively.  We then estimate the PDF of the radio sources in physical association with a $\gamma$-ray source as 
\[
P_{p} = \frac{a P_0 - b P_{\Delta\alpha}}{a-b}
\]
where $a+b = 1$ and $a/b = n_0/n_{\Delta\alpha}$.  For each candidate we can then compute the likelihood of it belonging to the distribution $P_{p}$.  Independently, the probability of a chance alignment can be estimated from the pair separation relative to $R_{95}$, the 95 per cent confidence radius quoted for each 2FGL source.  We use the combination of these two probabilities to rank the candidates.  This procedure is analogous to that used by \cite{Abdo:2010fx} in the association between the First Fermi-LAT catalogue (1FGL) and the CRATES radio catalogue.  

After ranking the list, we regard the top 127 as likely associations.  See  appendix \ref{atpmn2fglassociation} for a list of ATPMN--2FGL associations. In the publication of the 2FGL \citep{Collaboration:2011tta}, likely associations with many of the 2FGL sources are listed.  All but one of the likely 127 has an association suggested in the 2FGL.  The unassociated source is 2FGL J1353.5-6640 which lies 1.2 arcmin from ATPMN J135340.1-663957 which has a 4.8~GHz flux density of 43 mJy and a spectral index of $-0.8\pm0.7$.  Table \ref{fermimatches} summarises the ATPMN--2FGL associations.

ATPMN J135340.1-663957, the new association with 2FGL J1353.5-6640, corresponds to a source studied by \cite{Tsarevsky:2005}, VASC J1353-66, in a search for very active stars in the Galaxy. \citeauthor{Tsarevsky:2005} studied objects in the plane of Galactic that have both a hard X-ray and radio emission. The radio counterparts were observed with the VLA or the ATCA to determine accurate positions and allow optical identification ans spectral measurement.  VASC~J1353$-$66 was identified with a  $R$ magnitude 17.1 object with a featureless spectrum, suggesting that it is a BL Lac type object.  The ATCA measurements of the object (more sensitive than the ATPMN observations) gave flux densities at 4.8 and 8.6~GHz of $48.2\pm0.1$~mJy and $41.7\pm0.1$ mJy respectively, compared with the ATPMN determinations of $43\pm6$ and $28\pm10$ mJy.

\begin{table*}
 \centering
 \caption{Summary of ATPMN--2FGL associations.  The  330 2FGL sources in the ATPMN sky are classified according to the likelihood of association with an ATPMN source.  The categories of associated objects are as used in the 2FGL catalogue:  {\bf AG} : (agn,agu)  other non-blazar agn, active galaxy of uncertain type.  {\bf BZ} : (bzb,bzq) BL Lac type of blazar, FSRQ (flat spectrum radio quasar) type of blazar.   {\bf PSR} : pulsar, either with or without LAT-detected pulsations. {\bf rdg} : radio galaxy. {\bf sbg} : starburst galaxy.  {\bf sey} : Seyfert galaxy.   {\bf spp} : special case - potential association with SNR or PWN.}

\begin{tabular}{lrrrrrrrrrr}
  \hline
 Class  & Number & \multicolumn{9}{c}{Associations} \\
           &               & AG & BZ & PSR & rdg  & sbg & sey & spp & None & AT20G \\
\hline
Likely associations     & 127 &  58 & 65  &   0 & 1 & 1 & 1 & 0 & 1 & 67 \\
Unlikely associations  &  28 &    6 & 10  &  1 & 0 & 0 & 0 & 1  & 10 &  12 \\
No ATPMN source            & 175 &  25 & 11  & 12 & 0 & 0 & 1 & 5 & 121 &  5 \\
\end{tabular}
  \label{fermimatches}
\end{table*}

\subsection{Variability}
\label{variability}
\begin{figure}
\begin{center}
\putfig{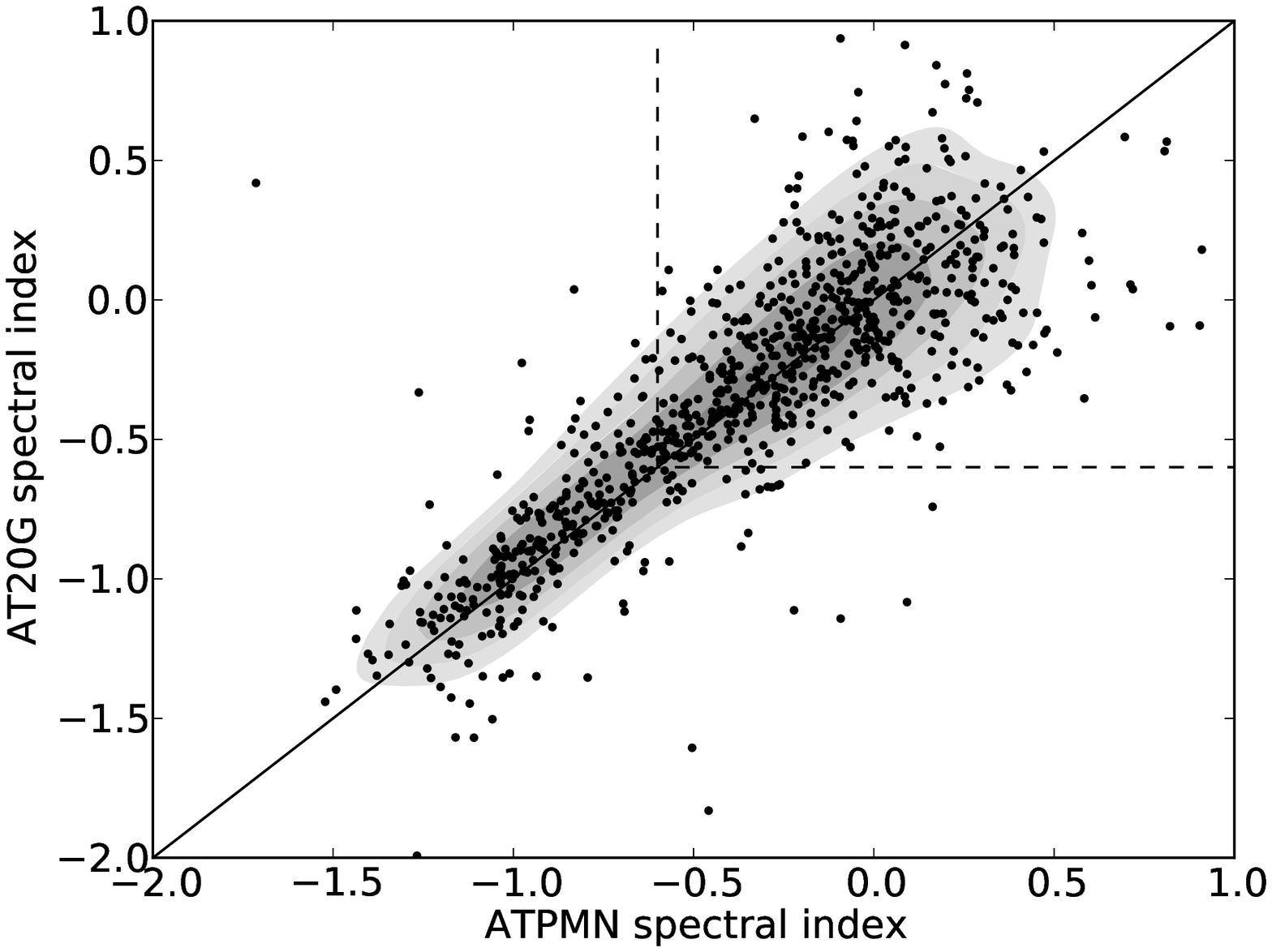}{\columnwidth}
\caption{Comparison of ATPMN and AT20G spectral indices, showing a tendency for greater variability when $\alpha \ga -0.6$.  The effect is less marked than that shown in Fig. \ref{atcalSI}: scatter is introduced from both the larger proportion of weaker sources and the different spatial frequency response of the two sets of observations.}
\label{at20gSI}
\end{center}
\end{figure}

In section \textsection\ref{radiometry} we noted, in the comparison of ATPMN calibrator observations with the data in the ATCA calibrator catalogue, that sources with $\alpha_{4.8}^{8.6} \ga -0.6$ have variable spectral indices. Also in that section we distinguish between  steep and flat-spectrum sources in the analysis of multiple observations of the same field during the ATPMN survey: steep-spectrum sources have more consistent flux density.  The same effect is evident in a comparison of ATPMN and AT20G source spectral indices.  These two surveys measured source flux densities at widely spaced epochs and, as shown in Fig. \ref{at20gSI}, the spectral index of those sources with steep spectra ($\alpha \la -0.6$) tends to be less variable.  These observations are consistent with the picture of radio sources in which those with flat spectra are identified with the bright compact cores of AGNs, and those with steep spectra with older populations of synchrotron-emitting particles, further from AGN cores, in radio lobes or elsewhere, and with larger angular size.

\subsection{Future analysis}
There are (at least) three aspects of the data yet to be analysed that will be the subject of future work.  The ATCA observations measured all components of the radiation, allowing a full polarimetric characterisation of each source.  Many of the fields bear evidence of extended emission that has not been characterised.  In the present analysis, preference has been given to characterising the more compact sources.  Different techniques will be needed to quantify the extended emission, particularly when it appears in fields that also contain compact sources.  Finally, visual examination of the visibility data suggests a class of field that contain two closely separated sources, but the automatic procedure used here interprets it a single resolved sources.  Again, some different techniques will be developed to adequately characterise such fields.

\section{Conclusions}

We present a new catalogue of \Nsources radio sources in the southern sky, with flux density, position and angular size measurements at 4.8 and 8.64 GHz.  The relatively high resolution of the observations allows position accuracy of typically 0\farcs4, as determined by comparison with the International Celestial Reference Frame.  This will allow improved cross-identification with observations in other wave bands, particularly with optical images.  The new catalogue increases our knowledge of the brighter PMN sources, allowing the classification of sources according to spectral index and structure.  The catalogue includes more than 2000 flat-spectrum ($\alpha^{8.6}_{4.8} > -0.5$) sources with 4.8~GHz flux density greater than 25mJy. It includes 2096 sources that lie with 5\arcsec of a source in the AT20G catalogue, and approximately 127 counterparts to $\gamma$-ray sources in the Fermi 2FGL catalogue.   We have made a new radio\,$-$\,$\gamma$-ray association between ATPMN J135340.1-663957 and 2FGL J1353.5-6640, a source previously studied by \cite{Tsarevsky:2005} as VASC J1353-66, and tentatively identified  them as a BL Lac type object.

Future work will process and publish the polarization properties of the ATPMN sources, and investigate several interesting classes of source. Sources that are the barely-resolved or closely-paired flat-spectrum sources that will be examined for evidence of gravitational lensing.  Sources that show signs of variability during the $\sim8$ hours that spanned the observation of each field will be studied as potential intra-day scintillators.  

\section*{Acknowledgements}

This paper includes archived data obtained through the Australia Telescope Online Archive (http://atoa.atnf.csiro.au).
It has also made use of data obtained from the SuperCOSMOS Science Archive, prepared and hosted by the Wide Field Astronomy Unit, Institute for Astronomy, University of Edinburgh, which is funded by the UK Science and Technology Facilities Council. 

\bibliography{dmbib,ATPMN}{}
\bibliographystyle{mn2e}

\appendix
\section{ATPMN-2FGL associations}
\label{atpmn2fglassociation}

In \textsection\ref{fermilat} we described the matching procedure followed to associate ATPMN sources with sources in the 2FGL.  We estimated there to be $N_{\gamma}\pm\sigma_N$ = $127\pm5$ real associations out of the 154 2FGL sources with at least one matching ATPMN source to within $1.4 \times R_{95}$.  In many cases several ATPMN sources lie inside the matching radius so that there is a total of 186 ATPMN sources with a position match to a 2FGL source.   We list the possible ATPMN associations in three tables: Table \ref{likelyfermi} gives the most likely $N_{\gamma} - \sigma_N = 122$ associations; Table \ref{possiblefermi} lists $2 \sigma_N = 10$ associations of intermediate likelihood; Table \ref{unlikelyfermi} lists the source pairs close on the sky, but unlikely to be real associations.  In cases of more than one ATPMN source appearing close to a 2FGL source, we have removed the less likely associations from the lists in Tables \ref{likelyfermi} and \ref{possiblefermi}, and listed the alternative ATPMN associations in Table \ref{multiplefermi}.

\begin{table}
 \centering
 \caption{High probability associations between sources in the 2FGL and ATPMN.  The associations lised as ``original'' are from the paper announcing the 2FGL results \citep{Collaboration:2011tta}. }
  \label{likelyfermi}
{\scriptsize
\begin{tabular}{lll}
 \hline
  \multicolumn{1}{c}{2FGL}  &   \multicolumn{2}{c}{Association}  \\
                                           &   \multicolumn{1}{c}{ATPMN}  &   \multicolumn{1}{c}{Original}  \\
  \hline
 J0004.7$-$4736 & J000435.6$-$473619 & PKS 0002$-$478 \\
 J0012.9$-$3954 & J001259.9$-$395426 & PKS 0010$-$401 \\
 J0018.8$-$8154 & J001920.6$-$815251 & PMN J0019$-$8152 \\
 J0029.2$-$7043 & J002841.6$-$704516 & PKS 0026$-$710 \\
 J0030.2$-$4223 & J003017.5$-$422446 & PKS 0027$-$426 \\
 J0046.7$-$8416 & J004426.7$-$842239 & PKS 0044$-$84 \\
 J0049.7$-$5738 & J004959.4$-$573827 & PKS 0047$-$579 \\
 J0059.7$-$5700 & J005846.5$-$565911 & PKS 0056$-$572 \\
 J0146.6$-$5206 & J014648.6$-$520233 & PKS 0144$-$522 \\
 J0158.0$-$4609 & J015751.1$-$461423 & PMN J0157$-$4614 \\
 J0158.3$-$3931 & J015838.1$-$393203 & PMN J0158$-$3932 \\
 J0207.9$-$6832 & J020750.9$-$683755 & PKS 0206$-$688 \\
 J0210.7$-$5102 & J021046.2$-$510102 & PKS 0208$-$512 \\
 J0237.1$-$6136 & J023653.2$-$613615 & PKS 0235$-$618 \\
 J0245.9$-$4652 & J024600.1$-$465117 & PKS 0244$-$470 \\
 J0302.7$-$7919 & J030320.9$-$791456 & PMN J0303$-$7914 \\
 J0303.5$-$6209 & J030350.6$-$621125 & PKS 0302$-$623 \\
 J0310.0$-$6058 & J030956.0$-$605839 & PKS 0308$-$611 \\
 J0325.1$-$5635 & J032523.5$-$563544 & 1RXS J032521.8$-$563543 \\
 J0334.2$-$4008 & J033413.6$-$400825 & PKS 0332$-$403 \\
 J0357.0$-$4950 & J035700.1$-$495548 & PKS 0355$-$500 \\
 J0424.3$-$5332 & J042504.2$-$533158 & PMN J0425$-$5331 \\
 J0433.4$-$6029 & J043334.1$-$603013 & PKS 0432$-$606 \\
 J0438.8$-$4521 & J043900.8$-$452222 & PKS 0437$-$454 \\
 J0449.4$-$4350 & J044924.7$-$435008 & PKS 0447$-$439 \\
 J0456.1$-$4613 & J045550.7$-$461558 & PKS 0454$-$46 \\
 J0507.5$-$6102 & J050754.6$-$610443 & PMN J0507$-$6104 \\
 J0516.5$-$4601 & J051545.2$-$455643 & PKS 0514$-$459 \\
 J0516.8$-$6207 & J051644.9$-$620705 & PKS 0516$-$621 \\
 J0526.1$-$4829 & J052616.6$-$483036 & PKS 0524$-$485 \\
 J0531.8$-$8324 & J053338.3$-$832435 & PKS 0541$-$834 \\
 J0532.0$-$4826 & J053158.6$-$482735 & PMN J0531$-$4827 \\
 J0532.5$-$7223 & J053344.7$-$721624 & PMN J0533$-$7216 \\
 J0538.8$-$4405 & J053850.3$-$440508 & PKS 0537$-$441 \\
 J0558.7$-$7501 & J055846.0$-$745905 & PKS 0600$-$749 \\
 J0601.1$-$7037 & J060111.2$-$703608 & PKS 0601$-$70 \\
 J0628.9$-$6246 & J062857.4$-$624845 & PKS 0628$-$627 \\
 J0635.5$-$7516 & J063546.4$-$751616 & PKS 0637$-$75 \\
 J0644.2$-$6713 & J064428.0$-$671257 & PKS 0644$-$671 \\
 J0647.8$-$6102 & J064740.8$-$605805 & PMN J0647$-$6058 \\
 J0700.3$-$6611 & J070031.2$-$661045 & PKS 0700$-$661 \\
 J0701.7$-$4630 & J070134.5$-$463436 & PKS 0700$-$465 \\
 J0706.7$-$4845 & J070558.7$-$484724 & PMN J0705$-$4847 \\
 J0718.7$-$4320 & J071843.6$-$431949 & PMN J0718$-$4319 \\
 J0727.0$-$4726 & J072626.2$-$472853 & PMN J0726$-$4728 \\
 J0730.6$-$6607 & J073049.5$-$660218 & PMN J0730$-$6602 \\
 J0734.2$-$7706 & J073443.4$-$771113 & PKS 0736$-$770 \\
 J0746.5$-$4758 & J074642.3$-$475455 & CRATES J0746$-$4754 \\
 J0844.8$-$5459 & J084502.4$-$545808 & PMN J0845$-$5458 \\
 J0852.4$-$5756 & J085238.7$-$575529 & PMN J0852$-$5755 \\
 J0904.9$-$5735 & J090453.2$-$573504 & PKS 0903$-$57 \\
 J0940.8$-$6105 & J094047.3$-$610726 & MRC 0939$-$608 \\
 J0942.8$-$7558 & J094509.2$-$755833 & PKS 0943$-$76 \\
 J1026.3$-$8546 & J102634.4$-$854314 & PKS 1029$-$85 \\
 J1047.7$-$6216 & J104742.9$-$621714 & PMN J1047$-$6217 \\
 J1103.9$-$5356 & J110352.1$-$535700 & PKS 1101$-$536 \\
 J1107.2$-$4448 & J110708.6$-$444907 & PKS 1104$-$445 \\
 J1117.2$-$4844 & J111719.9$-$483810 & CRATES J1117$-$4838 \\
 J1118.1$-$4629 & J111826.9$-$463414 & PKS 1116$-$46 \\
 J1134.4$-$7415 & J113609.7$-$741546 & PKS 1133$-$739 \\
 J1135.2$-$6829 & J113602.0$-$682705 & PKS 1133$-$681 \\
 J1218.8$-$4827 & J121902.2$-$482628 & PMN J1219$-$4826 \\
 J1228.7$-$8310 & J122454.4$-$831310 & PKS 1221$-$82 \\
 J1303.5$-$4622 & J130340.2$-$462102 & PMN J1303$-$4621 \\
 J1303.8$-$5537 & J130349.2$-$554031 & PMN J1303$-$5540 \\
 J1305.8$-$4925 & J130532.2$-$492829 & NGC 4945 \\
 J1307.6$-$6704 & J130817.3$-$670705 & PKS 1304$-$668 \\
 J1314.5$-$5330 & J131504.1$-$533435 & PMN J1315$-$5334 \\
 J1326.7$-$5254 & J132649.2$-$525623 & PMN J1326$-$5256 \\
 J1329.2$-$5608 & J132901.1$-$560802 & PMN J1329$-$5608 \\
 J1330.1$-$7002 & J133011.0$-$700313 & PKS 1326$-$697 \\
 J1352.6$-$4413 & J135256.5$-$441240 & PKS 1349$-$439 \\
 J1353.5$-$6640 & J135340.1$-$663957 & $-$ \\
 J1400.6$-$5601 & J140041.7$-$560455 & PMN J1400$-$5605 \\
 J1443.9$-$3908 & J144357.2$-$390840 & PKS 1440$-$389 \\
 J1508.5$-$4957 & J150838.9$-$495302 & PMN J1508$-$4953 \\

\hline

\end{tabular}
}
\end{table}

\begin{table}
 \centering
 \contcaption{High probability associations between sources in the 2FGL and ATPMN.}
{\scriptsize
\begin{tabular}{lll}
  \hline
  \multicolumn{1}{c}{2FGL}  &   \multicolumn{2}{c}{Association}  \\
                                           &   \multicolumn{1}{c}{ATPMN}  &   \multicolumn{1}{c}{Original}  \\
  \hline
 J1508.9$-$4342 & J150935.7$-$434031 & PMN J1509$-$4340 \\
 J1514.6$-$4751 & J151440.0$-$474829 & PMN J1514$-$4748 \\
 J1537.4$-$7957 & J153740.7$-$795805 & PMN J1537$-$7958 \\
 J1558.6$-$7039 & J155736.1$-$704027 & PKS 1552$-$705 \\
 J1558.9$-$6428 & J155850.3$-$643229 & PMN J1558$-$6432 \\
 J1603.8$-$4904 & J160350.6$-$490405 & PMN J1603$-$4904 \\
 J1604.5$-$4442 & J160431.0$-$444131 & PMN J1604$-$4441 \\
 J1610.6$-$4002 & J161021.8$-$395858 & PMN J1610$-$3958 \\
 J1610.8$-$6650 & J161046.4$-$664901 & PMN J1610$-$6649 \\
 J1618.2$-$7718 & J161749.1$-$771718 & PKS 1610$-$77 \\
 J1626.0$-$7636 & J162638.1$-$763855 & PKS 1619$-$765 \\
 J1650.1$-$5044 & J165016.6$-$504448 & PMN J1650$-$5044 \\
 J1725.1$-$7714 & J172350.8$-$771350 & PKS 1716$-$771 \\
 J1753.8$-$5012 & J175338.5$-$501514 & PMN J1753$-$5015 \\
 J1755.5$-$6423 & J175442.0$-$642345 & PMN J1754$-$6423 \\
 J1759.2$-$4819 & J175858.4$-$482112 & PMN J1758$-$4820 \\
 J1802.6$-$3940 & J180242.6$-$394007 & PMN J1802$-$3940 \\
 J1816.7$-$4942 & J181655.9$-$494344 & PMN J1816$-$4943 \\
 J1825.1$-$5231 & J182513.8$-$523058 & PKS 1821$-$525 \\
 J1830.2$-$4441 & J183000.8$-$444111 & PMN J1830$-$4441 \\
 J1832.7$-$5700 & J183230.9$-$565920 & PMN J1832$-$5659 \\
 J1849.7$-$4310 & J184925.8$-$431413 & PMN J1849$-$4314 \\
 J1902.5$-$6746 & J190301.2$-$674935 & PMN J1903$-$6749 \\
 J1918.2$-$4110 & J191816.0$-$411130 & PMN J1918$-$4111 \\
 J1937.2$-$3955 & J193716.2$-$395801 & PKS 1933$-$400 \\
 J1940.8$-$6213 & J194121.7$-$621121 & PKS 1936$-$623 \\
 J1959.1$-$4245 & J195913.2$-$424607 & PMN J1959$-$4246 \\
 J2009.5$-$4850 & J200925.3$-$484953 & PKS 2005$-$489 \\
 J2022.3$-$4518 & J202226.4$-$451329 & PMN J2022$-$4513 \\
 J2040.2$-$7109 & J204008.1$-$711459 & PKS 2035$-$714 \\
 J2056.2$-$4715 & J205616.3$-$471447 & PKS 2052$-$47 \\
 J2103.6$-$6236 & J210338.4$-$623225 & PMN J2103$-$6232 \\
 J2108.9$-$6636 & J210851.7$-$663722 & PKS 2104$-$668 \\
 J2134.5$-$6513 & J213413.2$-$651337 & PKS 2130$-$654 \\
 J2135.6$-$4959 & J213519.7$-$500647 & PMN J2135$-$5006 \\
 J2139.3$-$4236 & J213924.1$-$423520 & MH 2136$-$428 \\
 J2143.2$-$3929 & J214302.8$-$392924 & PMN J2143$-$3929 \\
 J2147.4$-$7534 & J214712.7$-$753613 & PKS 2142$-$75 \\
 J2201.9$-$8335 & J220219.1$-$833811 & PKS 2155$-$83 \\
 J2208.1$-$5345 & J220743.7$-$534633 & PKS 2204$-$54 \\
 J2234.9$-$4831 & J223513.2$-$483558 & PKS 2232$-$488 \\
 J2237.2$-$3920 & J223708.1$-$392138 & PKS 2234$-$396 \\
 J2250.2$-$4205 & J225022.2$-$420613 & PMN J2250$-$4206 \\
 J2319.1$-$4208 & J231905.8$-$420648 & PKS 2316$-$423 \\
 J2325.4$-$4758 & J232526.8$-$480017 & PKS 2322$-$482 \\
 J2329.2$-$4956 & J232920.8$-$495540 & PKS 2326$-$502 \\
\hline
\end{tabular}
}
\end{table}

\begin{table}
 \centering
 \caption{Possible associations between sources in the 2FGL and ATPMN. The associations lised as ``original'' are from the paper announcing the 2FGL results \citep{Collaboration:2011tta}. }
  \label{possiblefermi}
{\scriptsize
\begin{tabular}{lll}
  \hline
  \multicolumn{1}{c}{2FGL}  &   \multicolumn{2}{c}{Association}  \\
                                           &   \multicolumn{1}{c}{ATPMN}  &   \multicolumn{1}{c}{Original}  \\
  \hline
 J0413.5$-$5332 & J041313.4$-$533200 & PMN J0413$-$5332 \\
 J0540.4$-$5415 & J054045.8$-$541822 & PKS 0539$-$543 \\
 J0626.8$-$4258 & J062607.9$-$425332 & 1RXS J062635.9$-$42581 \\
 J0811.1$-$7527 & J081103.1$-$753027 & PMN J0810$-$7530 \\
 J1057.0$-$8004 & J105843.2$-$800354 & PKS 1057$-$79 \\
 J1230.2$-$5258 & J122939.8$-$530332 & PMN J1229$-$5303 \\
 J2007.9$-$4430 & J200755.1$-$443444 & PKS 2004$-$447 \\
 J2315.7$-$5014 & J231544.3$-$501839 & PKS 2312$-$505 \\
 J2327.9$-$4037 & J232819.2$-$403509 & PKS 2325$-$408 \\
 J2329.7$-$4744 & J232917.7$-$473019 & PKS 2326$-$477 \\
 J2336.3$-$4111 & J233633.9$-$411522 & PKS 2333$-$415 \\

\hline
\end{tabular}
}
\end{table}

\begin{table}
 \centering
 \caption{Unlikely associations between sources in the 2FGL and ATPMN. The associations lised as ``original'' are from the paper announcing the 2FGL results \citep{Collaboration:2011tta}. }
  \label{unlikelyfermi}
{\scriptsize
\begin{tabular}{lll}
  \hline
  \multicolumn{1}{c}{2FGL}  &   \multicolumn{2}{c}{Association}  \\
                                           &   \multicolumn{1}{c}{ATPMN}  &   \multicolumn{1}{c}{Original}  \\
  \hline
 J0022.3$-$5141 & J002233.3$-$515316 & 1RXS J002159.2$-$514028 \\
 J0029.2$-$7043 & J003120.4$-$703649 & PKS 0026$-$710 \\
 J0046.7$-$8416 & J004425.4$-$841438 & PKS 0044$-$84 \\
 J0201.5$-$6626 & J020107.7$-$663812 & PMN J0201$-$6638 \\
 J0216.9$-$6630 & J021650.7$-$663643 & RBS 0300 \\
 J0216.9$-$6630 & J021656.9$-$663646 & RBS 0300 \\
 J0310.2$-$5013 & J031033.5$-$501005 & SUMSS J031034$-$501633 \\
 J0316.1$-$6434 & J031513.9$-$643636 & $-$ \\
 J0316.1$-$6434 & J031515.1$-$643508 & $-$ \\
 J0325.1$-$5635 & J032522.6$-$562905 & 1RXS J032521.8$-$563543 \\
 J0416.0$-$4355 & J041643.5$-$440217 & $-$ \\
 J0456.1$-$4613 & J045550.4$-$461555 & PKS 0454$-$46 \\
 J0532.5$-$7223 & J053254.0$-$723155 & PMN J0533$-$7216 \\
 J0532.5$-$7223 & J053254.3$-$723154 & PMN J0533$-$7216 \\
 J0611.8$-$6059 & J061030.2$-$605838 & PKS 0609$-$609 \\
 J0821.0$-$4254 & J082035.2$-$430023 & $-$ \\
 J0821.0$-$4254 & J082035.6$-$430022 & $-$ \\
 J0942.8$-$7558 & J094507.7$-$755749 & PKS 0943$-$76 \\
 J1118.1$-$4629 & J111826.7$-$463414 & PKS 1116$-$46 \\
 J1134.4$-$7415 & J113219.1$-$742508 & PKS 1133$-$739 \\
 J1228.7$-$8310 & J123109.1$-$831446 & PKS 1221$-$82 \\
 J1228.7$-$8310 & J123112.7$-$831451 & PKS 1221$-$82 \\
 J1305.8$-$4925 & J130527.5$-$492805 & NGC 4945 \\
 J1307.5$-$4300 & J130737.9$-$425938 & 1RXS J130737.8$-$425940 \\
 J1312.0$-$6458 & J130400.3$-$645644 & $-$ \\
 J1312.0$-$6458 & J130408.2$-$645641 & $-$ \\
 J1325.6$-$4300 & J132527.5$-$430108 & Cen A Core \\
 J1407.5$-$4257 & J140739.7$-$430231 & CGRaBS J1407$-$4302 \\
 J1428.0$-$4206 & J142756.2$-$420619 & PKS B1424$-$418 \\
 J1437.2$-$5211 & J143815.2$-$521745 & $-$ \\
 J1539.3$-$4636 & J153938.3$-$464323 & $-$ \\
 J1610.1$-$4808 & J161056.5$-$480152 & $-$ \\
 J1610.1$-$4808 & J161113.1$-$480345 & $-$ \\
 J1629.6$-$6141 & J162854.7$-$615236 & PMN J1629$-$6133 \\
 J1629.6$-$6141 & J162909.3$-$613345 & PMN J1629$-$6133 \\
 J1658.4$-$5322 & J165909.0$-$532016 & PSR J1658$-$5324 \\
 J1658.4$-$5322 & J165909.5$-$532026 & PSR J1658$-$5324 \\
 J1958.2$-$3848 & J195759.8$-$384506 & PKS 1954$-$388 \\
 J2040.2$-$7109 & J204007.9$-$711453 & PKS 2035$-$714 \\
 J2040.2$-$7109 & J204011.0$-$711500 & PKS 2035$-$714 \\
 J2131.0$-$5417 & J212959.2$-$541354 & $-$ \\
 J2131.0$-$5417 & J213208.3$-$542036 & $-$ \\
 J2135.6$-$4959 & J213520.1$-$500651 & PMN J2135$-$5006 \\
 J2259.0$-$8254 & J225759.5$-$824653 & $-$ \\

\hline
\end{tabular}
}
\end{table}

\begin{table}
 \centering
 \caption{2FGL sources that lies within an angular distance $1.4 R_{95}$ of more than one ATPMN sources. }
  \label{multiplefermi}
{\scriptsize
\begin{tabular}{ll}
  \hline
  \multicolumn{1}{c}{2FGL}  &   \multicolumn{1}{c}{ATPMN}  \\
                                           &   \multicolumn{1}{c}{associations} \\
  \hline
 J0904.9$-$5735 & J090453.2$-$573504 \\
                            & J090453.3$-$573501 \\
                            & J090453.4$-$573503 \\
 J1134.4$-$7415 & J113609.6$-$741545 \\
                            & J113609.7$-$741546 \\
 J1303.8$-$5537 & J130349.2$-$554031 \\
                            & J130354.7$-$554100 \\
 J1508.5$-$4957 & J150838.9$-$495302 \\
                            & J150935.7$-$434031 \\
 J1825.1$-$5231 & J182511.5$-$523004 \\
                            & J182513.8$-$523058 \\
 J2208.1$-$5345 & J220743.4$-$534639 \\
                            & J220743.7$-$534633  \\
 J2237.2$-$3920 & J223704.7$-$391932 \\
                            & J223708.1$-$392138 \\

\hline
\end{tabular}
}
\end{table}

\label{lastpage}
\begin{onecolumn}
\begin{sidewaystable}\label{t_main}
{\bf Table 5.} A sample of 30 sources in the ATPMN catalogue. The columns are defined in \textsection\ref{theCatalogue}.\\
\label{samplecatalogue}
\setlength{\tabcolsep}{1mm}
 \begin{tabular}{@{}llllrrrrrrrrrrrrcc}
  \hline
     \multicolumn{1}{c}{IAU designation} & \multicolumn{1}{c}{PMN source} 
     & \multicolumn{1}{c}{$\alpha$} &  \multicolumn{1}{c}{$\delta$}
     &\multicolumn{1}{c}{$S_5$}   & \multicolumn{1}{c}{$\sigma_e$}
    & \multicolumn{1}{c}{$S_8$} & \multicolumn{1}{c}{$\sigma_8$} & \multicolumn{1}{c}{Maj$_5$} &\multicolumn{1}{c}{Min$_5$}
     &\multicolumn{1}{c}{PA$_5$} & \multicolumn{1}{c}{Maj$_8$} &\multicolumn{1}{c}{Min$_8$} &\multicolumn{1}{c}{PA$_8$}
     &\multicolumn{1}{c}{$\alpha$} &   \multicolumn{1}{c}{$\sigma_{\alpha}$} &  \multicolumn{1}{c}{Ep}   &\multicolumn{1}{c}{QFlg} \\
         & & \multicolumn{1}{c}{(J2000)} & \multicolumn{1}{c}{(J2000)} 
     &\multicolumn{2}{c}{(mJy)}  
    & \multicolumn{2}{c}{(mJy)} & \multicolumn{2}{c}{(\arcsec)}
     & \multicolumn{1}{c}{(\degr)} & \multicolumn{2}{c}{(\arcsec)} &\multicolumn{1}{c}{(\degr)}
     & &   &  & \\

\hline
J161113.1$-$480345 & PMNJ1611$-$4802  & 16:11:13.16  & $-$48:03:45.1   &   46 &   7 &   12 &  10 &    0.7 &  0.6 &  23 &    0.0 &  0.0 &   0 & $-$2.2 & 1.4 &   4 & 0 0 \\
J161144.8$-$601206 & PMNJ1611$-$6012  & 16:11:44.89  & $-$60:12:06.7   &   80 &   7 &   42 &  10 &    0.0 &  0.0 &   0 &    0.0 &  0.0 &   0 & $-$1.1 & 0.4 &   1 & 0 1 \\
J161147.4$-$550855 & PMNJ1611$-$5508  & 16:11:47.40  & $-$55:08:55     &   50 &   7 &   25 &  10 &    1.8 &  0.9 & $-$80 &    2.4 &  0.8 &  32 & $-$1.2 & 0.7 &   3 & 0 0 \\
J161148.8$-$550901 & PMNJ1611$-$5508  & 16:11:48.81  & $-$55:09:01     &   29 &   7 &   17 &  10 &    1.2 &  0.2 & $-$33 &    1.7 &  0.1 & $-$30 & $-$0.9 & 1.1 &   3 & 0 0 \\
J161200.2$-$470137 & PMNJ1611$-$4701  & 16:12:00.27  & $-$47:01:37.6   &   48 &   7 &   16 &  10 &    2.5 &  0.7 &  35 &    0.9 &  0.5 & $-$60 & $-$1.9 & 1.1 &   4 & 0 0 \\
\\
J161233.7$-$742340 & PMNJ1612$-$7423  & 16:12:33.77  & $-$74:23:40.1   &  167 &   7 &  220 &  14 &    0.0 &  0.0 &   0 &    0.0 &  0.0 &   0 &  0.5 & 0.1 &   1 & 0 0 \\
J161242.8$-$390644 & PMNJ1612$-$3906  & 16:12:42.8   & $-$39:06:44     &   86 &   7 &   38 &  10 &    5.7 &  2.9 & $-$55 &    4.6 &  2.1 & $-$51 & $-$1.4 & 0.5 &   5 & 0 0 \\
J161256.2$-$795834 & PMNJ1613$-$7958  & 16:12:56.2   & $-$79:58:34     &   24 &   7 &   14 &  10 &    1.7 &  0.5 &  36 &    2.1 &  0.1 &  88 & $-$1.0 & 1.3 &   1 & 0 0 \\
J161256.8$-$573647 & PMNJ1612$-$5736  & 16:12:56.8   & $-$57:36:47.2   &   45 &   7 &   30 &  10 &    2.0 &  0.7 & $-$75 &    1.6 &  0.1 & $-$12 & $-$0.7 & 0.6 &  12 & 0 1 \\
J161316.9$-$480635 & PMNJ1613$-$4806  & 16:13:16.93  & $-$48:06:35.7   &  105 &   7 &   46 &  10 &    1.1 &  0.2 &  32 &    0.8 &  0.4 &  24 & $-$1.4 & 0.4 &   4 & 0 0 \\
\\
J161316.9$-$480650 & PMNJ1613$-$4806  & 16:13:16.98  & $-$48:06:50.8   &   54 &   7 & $-$ & $-$ &    3.2 &  0.2 &  $-$7 & $-$ & $-$ & $-$ & $-$ & $-$ &   4 & 0 0 \\
J161339.7$-$422744 & PMNJ1613$-$4225  & 16:13:39.7   & $-$42:27:44     &   44 &   7 &   27 &  10 &    4.9 &  0.2 &  19 &    3.0 &  0.5 &  27 & $-$0.8 & 0.7 &   5 & 1 0 \\
J161344.0$-$422540 & PMNJ1613$-$4225  & 16:13:44.02  & $-$42:25:40.3   &  293 &   8 &  165 &  12 &    0.0 &  0.0 &   0 &    0.7 &  0.1 &  $-$2 & $-$1.0 & 0.1 &   5 & 1 0 \\
J161408.8$-$791800 & PMNJ1614$-$7918  & 16:14:08.84  & $-$79:18:00.2   &   58 &   7 &   81 &  11 &    0.0 &  0.0 &   0 &    0.0 &  0.0 &   0 &  0.6 & 0.3 &   1 & 0 0 \\
J161430.6$-$562241 & PMNJ1614$-$5622  & 16:14:30.6   & $-$56:22:41     &  122 &   7 &   69 &  10 &    8.2 &  0.7 & $-$80 &    4.6 &  0.1 & $-$75 & $-$1.0 & 0.3 &   1 & 0 1 \\
\\
J161431.2$-$562246 & PMNJ1614$-$5622  & 16:14:31.25  & $-$56:22:46.6   &  501 &  10 &  308 &  17 &    1.0 &  0.2 &  90 &    0.4 &  0.2 &  54 & $-$0.8 & 0.1 &   1 & 0 1 \\
J161433.3$-$562245 & PMNJ1614$-$5622  & 16:14:33.3   & $-$56:22:45     &  115 &   7 &   63 &  10 &    4.4 &  1.1 & $-$66 &    3.6 &  1.0 & $-$68 & $-$1.0 & 0.3 &   1 & 0 1 \\
J161433.5$-$421425 & PMNJ1614$-$4214  & 16:14:33.5   & $-$42:14:25.9   &   46 &   7 &   28 &  10 &    3.1 &  0.8 & $-$79 &    2.2 &  0.1 &  $-$9 & $-$0.8 & 0.7 &   5 & 0 0 \\
J161441.9$-$442449 & PMNJ1614$-$4424  & 16:14:41.96  & $-$44:24:49.0   &  168 &   7 &   70 &  10 &    2.3 &  0.6 &  29 &    1.9 &  0.1 &  33 & $-$1.5 & 0.3 &   4 & 0 0 \\
J161442.2$-$442442 & PMNJ1614$-$4424  & 16:14:42.29  & $-$44:24:42.0   &  138 &   7 &   73 &  11 &    1.1 &  0.9 &  21 &    0.6 &  0.1 &  $-$5 & $-$1.1 & 0.3 &   4 & 0 0 \\
\\
J161442.4$-$442439 & PMNJ1614$-$4424  & 16:14:42.41  & $-$44:24:39.3   &  173 &   7 &  103 &  11 &    0.7 &  0.2 & $-$67 &    0.4 &  0.1 &  79 & $-$0.9 & 0.2 &   4 & 0 0 \\
J161445.1$-$571910 & PMNJ1614$-$5719  & 16:14:45.14  & $-$57:19:10.3   &  104 &   7 &  116 &  11 &    0.8 &  0.3 & $-$81 &    0.0 &  0.0 &   0 &  0.2 & 0.2 &  12 & 0 1 \\
J161451.6$-$434122 & PMNJ1614$-$4341  & 16:14:51.62  & $-$43:41:22.3   &   63 &   7 &   38 &  10 &    0.8 &  0.2 &  50 &    0.4 &  0.1 &  13 & $-$0.9 & 0.5 &   4 & 0 0 \\
J161456.8$-$590514 & PMNJ1615$-$5905  & 16:14:56.8   & $-$59:05:14     &   31 &   7 &   20 &  10 &    1.9 &  0.2 & $-$78 &    2.6 &  0.1 & $-$72 & $-$0.8 & 0.9 &  12 & 0 0 \\
J161502.1$-$590601 & PMNJ1615$-$5905  & 16:15:02.1   & $-$59:06:01.2   &   20 &   7 &   12 &  10 &    1.5 &  0.2 &   2 &    1.5 &  0.1 & $-$81 & $-$0.9 & 1.5 &  12 & 0 0 \\
\\
J161502.3$-$590552 & PMNJ1615$-$5905  & 16:15:02.37  & $-$59:05:52.3   &   27 &   7 &   11 &  10 &    1.3 &  0.4 &  14 &    0.7 &  0.3 & $-$75 & $-$1.5 & 1.6 &  12 & 0 0 \\
J161502.8$-$451154 & PMNJ1615$-$4511  & 16:15:02.82  & $-$45:11:54.0   &   62 &   7 & $-$ & $-$ &    5.2 &  4.0 & $-$66 & $-$ & $-$ & $-$ & $-$ & $-$ &   4 & 0 0 \\
J161503.8$-$605426 & PMNJ1615$-$6054  & 16:15:03.8   & $-$60:54:26.1   &  167 &   7 &  148 &  12 &    1.4 &  1.1 & $-$20 &    2.3 &  0.1 & $-$69 & $-$0.2 & 0.2 &  12 & 1 0 \\
J161505.3$-$405755 & PMNJ1615$-$4058  & 16:15:05.3   & $-$40:57:55     &   78 &   7 &   53 &  10 &    4.1 &  2.4 &  61 &    6.2 &  0.1 &  58 & $-$0.6 & 0.4 &   5 & 0 0 \\
J161506.8$-$451934 & PMNJ1615$-$4519  & 16:15:06.84  & $-$45:19:34.3   &   82 &   7 & $-$ & $-$ &   14.6 &  0.2 &  33 & $-$ & $-$ & $-$ & $-$ & $-$ &   4 & 0 0 \\

\hline
\end{tabular}
\flushleft
Notes:\\
\end{sidewaystable}
\end{onecolumn}

\label{lastpage}
\end{document}